\documentclass[12pt]{article}

\usepackage{graphicx} % Required for inserting images

\usepackage[numbers]{natbib}
\usepackage{caption}
\usepackage{booktabs}
\usepackage{tabularx}
\usepackage[utf8]{inputenc}
\usepackage{threeparttable}
\usepackage{lscape}
\usepackage{float}
\usepackage{mathtools}
\usepackage{longtable}
\usepackage{amsfonts}
\usepackage{amsmath}
\usepackage{mathabx}
\usepackage{graphicx}
\usepackage{bm}
\usepackage{xcolor}
\usepackage{appendix}
\usepackage[normalem]{ulem}
\usepackage{relsize}
\usepackage{adjustbox}
\usepackage{setspace}
\usepackage{enumitem}
\usepackage{tabularray}
\usepackage[margin=1in]{geometry}

\let\P\undefined
\DeclareMathOperator*{\P}{\mathbb{P}}

\title{A Non-Parametric Sensitivity Analysis for Bounding Bias in Hybrid Control Trials}
\author{\makebox[.9\textwidth]{Alissa Gordon}\\Division of Biostatistics\\ UC Berkeley  \and Emilie H\o jbjerre-Frandsen\\Biostatistics Methods, Novo Nordisk A/S \& \\ Department of Mathematical Sciences, Aalborg University \and Alejandro Schuler\\Division of Biostatistics\\ UC Berkeley
}
\date{July 2025}

\begin{document}

\maketitle

\begin{abstract}
    % In the digital era it is easier than ever to collect and exploit rich covariate information in trials. Recent work explores how to use this information to integrate external controls, including the use of hybrid control trials (HCTs) where a randomized controlled trial is augmented with external controls. HCTs are of particular interest due to their ability to preserve partial randomization while also improving trial efficiency. However, most HCT estimators rely on an unrealistic assumption: that the external controls are drawn from the same population as the trial subjects (perhaps conditionally on covariates). There has been little formal work to quantify the inevitable bias introduced from a violation of this assumption, slowing the acceptance of HCT designs. To address this, we introduce a non-parametric sensitivity analysis that recognizes that the assumption can be reframed as a ‘no unobserved confounders’ assumption. We leverage omitted variable bias methodologies to estimate the maximum bias introduced from unmeasured covariates, allowing for a critical evaluation of the causal gap that invalidates significant findings. We show this method reliably bounds bias while also allowing for gains in efficiency, contingent only on conservatively specifying the added explanatory power of the unmeasured confounder to the outcome and trial enrollment probability. We conclude by discussing considerations for designing and evaluating HCTs, drawing on insights from simulations and theoretical analyses.
We study hybrid control trials (HCTs), in which a randomized controlled trial (RCT) is augmented with external control patients. Existing approaches for HCTs typically assume conditional exchangeability of the concurrent and external controls to identify trial-specific effects. When violated, this can induce substantial unquantified bias, which in turn limits the acceptability of HCTs in regulatory settings. We treat violations of mean exchangeability as omitted variable bias and develop a non-parametric sensitivity analysis that (i) applies to the efficient, doubly robust HCT estimator of the trial-specific ATE, and (ii) delivers sharp bounds on the bias induced by unmeasured covariates. Building on recent work in double machine learning, our approach characterizes the maximal bias in terms of two partial R-squared sensitivity parameters: the additional explanatory power that unmeasured confounders could have for the outcome regression and for trial participation. For any given choice of these parameters, we construct valid confidence bounds for bias-adjusted treatment effects and visualize critical causal gaps via contour plots and robustness values that show how strong unmeasured confounding would need to be to overturn nominally significant HCT findings. Through simulations, we show that the method (i) reliably upper-bounds true bias, (ii) restores type I error control in settings where naïve HCT analysis is anti-conservative, and (iii) can still deliver meaningful power gains and RCT sample-size reductions even under moderate violations of mean exchangeability. We illustrate the approach in a phase III trial on diabetes, supplemented with external controls. We discuss practical guidelines for designing and evaluating HCTs, including external-data selection, sample-size allocation, and interpretation of sensitivity contours.
\end{abstract}

\section{Introduction}
\label{sec1}

Randomized controlled trials (RCTs) are widely regarded as the gold standard in causal inference because they eliminate confounding, allowing for unbiased estimation \cite{sox_methods_2012}. However, RCTs face many limitations, including high costs, ethical constraints, and recruitment difficulties, that make them infeasible or impossible at times \cite{bentley_conducting_2019,glennerster_chapter_2017,temple_placebo-controlled_2000}.  These challenges can compromise research quality through underpowering of trials. Such problems are particularly common in research areas such as rare diseases or vulnerable populations, including pregnant mothers and marginalized populations, as well as in Phase II trials \cite{noauthor_rare_2019,jahanshahi_use_2021}.

Hybrid control trials (HCTs) are an alternative to randomized controlled trials in which an RCT is supplemented with additional control data from external sources (also known as “pooling”). Although not common in practice, the use of HCTs has been proposed as early as 1976 but has since received attention due to their potential to alleviate some of the restrictions mentioned above on RCTs \cite{pocock_combination_1976, tan_augmenting_2022,baumfeld_andre_trial_2020}. Regulatory agencies, including the FDA and EMA, have shown a growing interest and commitment to real-world evidence from these trials \cite{noauthor_rare_2019,pocock_combination_1976,baumfeld_andre_trial_2020,noauthor_international_2000, noauthor_fda_2019, noauthor_fda_2019-2, noauthor_framework_2018}. The 21st Century Cures Act likewise acknowledges the need to accelerate medical product development in the face of lengthy randomized controlled trials \cite{noauthor_21st_2017}. `Borrowing’ information from external sources aligns with these goals and can decrease trial timelines and sample sizes. Furthermore, data availability continues to grow as trials are conducted and pharmaceutical companies participate in data sharing of concurrent control data. Using external controls also offers the opportunity to change the randomization ratio in the RCT portion of the HCT \cite{lim_minimizing_2018}. This allows for more patients to receive the treatment rather than the control, making the trial both more ethical and attractive to prospective participants. With the growth of data, increased support from regulatory agencies, and the existing challenges of traditional RCTs, the time has come to explore data borrowing alternatives such as hybrid control trials.

Multiple data borrowing methods have been proposed with different positions on the bias-variance tradeoff spectrum. At one end of the spectrum, non-HCT methods such as prognostic covariate adjustment train models on previous data to create prognostic scores for trial participants \cite{schuler_increasing_2022, liao_prognostic_2025}. These do not directly incorporate external data as additional control data but instead use the predictions as an additional covariate to adjust on. Thus, bias is not introduced with the use of the external data. However, since information is borrowed conservatively, the variance reductions are relatively small compared to other methods.

At the other end of the spectrum, the most drastic borrowing method is the single arm trial where trial participants receive only the treatment, and the control arm consists solely of external data. Note that this is not a form of an HCT as the control arm is not hybrid. Single-arm trials are only acceptable in rare cases since they rely fully upon external data to make comparisons of treatment against control, making them the most susceptible to bias from the use of external data \cite{noauthor_framework_2018, agrawal_use_2023}. 

HCT methodologies lie in the middle of the spectrum with their positions determined by how they incorporate the external data. Some HCT methods incorporate external data more directly by `pooling' external and trial control data together. By treating external data as trial data, researchers artificially increase the RCT sample size. Typically, variance decreases and power increases with an increase in sample size. However, the risk of bias is also substantially increased if and when the external population does not truly reflect the trial population. These methods are often referred to as ``static'' borrowing: gains in efficiency are highest, but robustness against bias is lowest. With static borrowing, researchers always pool the data, assuming that the external data is compatible \cite{viele_use_2014}. The artificial trial sample size increases accordingly to the size of the external sample size, rapidly improving trial power. However, assuming compatibility of the data (which is often unrealistic) introduces estimation bias and type I error inflation.

Besides these ``static'' borrowing approaches, researchers have also considered ``dynamic'' approaches \cite{tan_augmenting_2022}. Dynamic borrowing methods fall between the extremes of the bias-variance spectrum, assessing compatibility of control data to adaptively change reliance on external data. These include Bayesian approaches such as power priors and the commensurate prior model \cite{tan_augmenting_2022, duan_evaluating_2006, neuenschwander_note_2009, hobbs_hierarchical_2011}.  Bayesian power prior methods rely on a weighting power parameter that ranges from 0 (equivalent to an RCT with no external data) to 1 (equivalent to a fully pooled HCT where external and trial data are treated equally). This power parameter is given a prior distribution that then is updated based on the utilized data to optimize weighting of data. These approaches are attractive since they measure differences in the data while still cleverly incorporating external control data. However, large power gains generally cannot be realized without sacrificing type I error control \cite{koppschneider_power_2020, ghadessi_roadmap_2020}.

Frequentist dynamic borrowing includes the test-then-pool approach where data is first assessed for significant differences such as by hypothesis tests \cite{tan_augmenting_2022}. If differences are significant, no pooling occurs; otherwise, pooling proceeds. By testing beforehand, researchers attempt to minimize bias by only pooling when data incompatibility is statistically insignificant. However, if no pooling occurs, researchers solely rely on RCT data with no avenues for gains in efficiency. If pooling occurs, the data are treated as if they are additional trial control units, though still subject to some weighting imposed by the estimator. This can produce substantial power gains. However, this approach is criticized for its inability to consistently detect biases, leaving it vulnerable to bias from pooling \cite{li_revisit_2020}. Finally, more rigorous frequentist dynamic borrowing methods have been proposed that rely on adaptive lasso penalization to select a subset of suitable external controls \cite{gao_improving_2024}. Only external controls whose bias parameters are (or close to) 0 are utilized, resulting in minimized bias from pooling. This method works well compared to others, notably in terms of mean-squared error, especially when differences in control groups are large. However, issues with type I error control arise with finite samples, and gains in power are not as large as other pooling methods due to the conservative nature of selecting suitable controls when differences are already small. 

Due to the large potential to increase power, we focus on a method that allows for pooled control data. The trial-specific average treatment effect (ATE) is the primary parameter of interest in many data-borrowing experiments, and thus many estimators have been proposed in HCT literature \cite{valancius_causal_2024, li_improving_2023, laan_targeted_2011}. However, these HCT estimators rely on an unrealistic assumption: mean exchangeability of the controls \cite{li_improving_2023}. 
A violation of this assumption introduces bias into estimation, potentially inflating type I error rates or decreasing power depending on the direction of the bias. In the context of clinical trials --- and particularly of phase III confirmatory trials --- these errors are especially problematic. Literature has focused on the development of these estimators, but few, if any, have discussed a formal approach to quantify the bias introduced by an inevitable violation of this assumption, slowing the acceptance of HCTs \cite{valancius_causal_2024, noauthor_international_2000}. 

\subsection{Contributions}

To mitigate against this bias, we propose that stringent sensitivity analysis has the potential to maintain the reliability of findings from these trials. We extend research from \citet{chernozhukov_long_2024} where a non-parametric sensitivity analysis leverages efficient debiased machine learning and produces easily interpretable bounds on omitted variable bias in observational studies. In theory, the mean exchangeability assumption is met when all relevant covariates are controlled for confounding. It follows that a violation of the mean exchangeability assumption is a result of omitted variable bias. We modify this sensitivity analysis in the context of HCTs to reliably bound bias given plausible increases in the explanatory power from the omitted variables on the ``Riesz representer'' and outcome regression of the trial-specific ATE \cite{chernozhukov_long_2024}. In doing so, we examine robustness of results against user-supplied sensitivity parameters, which can be evaluated for plausibility with observed variables, instead of relying on the untestable mean exchangeability assumption for inference. Results of the sensitivity analysis can then be compared against the strength of evidence from HCTs to determine whether conclusions are still significant even after accounting for the potential bias.

The aim of this research is to expand upon the current literature exploring the practicality of HCTs in the clinical setting. By proposing a reliable assumption-free sensitivity analysis for the pooled estimator, we add credibility to results from HCTs, unlocking the benefits they have to offer. We believe that increasing acceptance of these trial designs will accelerate the development of new therapies, particularly benefiting under-funded or underrepresented areas of research.

\paragraph{Outline.} 
In Section \ref{sec2} we begin by defining necessary notation and preliminaries for the hybrid control trial setting. We also explore a doubly robust efficient estimator for the trial-specific average treatment effect \cite{li_improving_2023}. In Section \ref{sec3} we validate that it is possible to extend the omitted variable bias sensitivity analysis to the HCT setting. Through identification of the Riesz representer and outcome regression of the trial-specific ATE, we derive an estimator for the bias bound $B$. In Section \ref{sec4}, we use simulations to demonstrate that the risk of bias and error is minimized with the sensitivity analysis and that trial efficiency can be improved even when bias is present. We do this by showing (i.) that the true bias is reliably bounded by $B$, (ii.) that the sensitivity analysis corrects inflated type I error rates in HCTs, and (iii.) that power can be increased when using the HCT estimator combined with the sensitivity analysis. In Section \ref{sec5}, we provide a case study using data from prior clinical trials conducted by Novo Nordisk (AS). We demonstrate how to apply the sensitivity analysis to results from an HCT study and discuss important practical considerations. In Section \ref{sec6}, we summarize key findings and explore guidelines for designing and evaluating HCTs based on these results as well as the potential impacts of this research.

\section{Notation and Preliminaries}
\label{sec2}

\subsection{Notation}
Consider a hypothetical hybrid control trial, consisting of a randomized controlled trial and an external control arm. Let the variable $D$ denote trial participation with $D=1$ indicating that a subject was a participant in the randomized controlled trial and $D=0$ indicating that a subject was not a participant in the randomized controlled trial. Equivalently, $D=0$ implies the participant is part of the external control arm. Let the variable $Y$ represent the outcome of interest in the study and $X$ represent the measured covariate(s) in the study. Let $A$ represent the binary treatment of interest with $A=1$ indicating that a subject received the treatment of interest and $A=0$ indicating that a subject did not receive the treatment. Note that since the external data only consist of control data, a subject with $D=0$ implies $A=0$. The observed data follow structure $O=(D,A,X,Y)$. Furthermore, we assume that the data $O_1, ..., O_n$ are i.i.d. distributed from a true data distributing distribution $\P_0$.

\begin{table}[ht]
\centering
% @{}…@{} suppresses the default padding so your rules line up
\begin{tabularx}{\textwidth}{@{}lX@{}}
\toprule
Variable & Definition \\
\midrule
$D$   & Trial participation indicator: $D=1$ if in RCT, $D=0$ if external control. \\
$A$   & Treatment indicator: $A=1$ if treated, $A=0$ otherwise. \\
$X$   & Measured covariate(s). \\
$Y$   & Observed outcome, with $Y = A\,Y_1 + (1-A)\,Y_0$. \\
$O$   & Observed data structure $(D,A,X,Y)$. \\
\midrule
$Y_1$ & Counterfactual outcome under treatment ($a=1$). \\
$Y_0$ & Counterfactual outcome under control ($a=0$). \\
$Z$   & Unobserved confounders (s.t.\ $Y_0\perp\!\!\perp D\mid X,Z$). \\
$O^*$ & Full (latent) data structure $(D,A,X,Z,Y_1,Y_0)$. \\
\bottomrule
\end{tabularx}
\caption{Random variables and their definitions.}
\end{table}

We also consider the unobserved full dataset. Let $Y_1$ and $Y_0$ denote the counterfactual outcomes where $Y_a$ represents the outcome that would be observed under treatment $a \in \{0,1\}$. We define the observed outcome $Y$ under treatment $A$ with the counterfactual outcomes $Y_a$ such that $Y=A\cdot Y_1+(1-A)\cdot Y_0$ (others call this the ``causal consistency'' assumption). Assume that there is a set of unobserved random variables Z such that $Y_0\perp\kern-9pt\perp D |X,Z$. These represent our ``unobserved confounders" without which we cannot guarantee mean exchangeability. The full data structure follows as $O^*=(D, A, X, Z, Y_1, Y_0)$, which is i.i.d. distributed from some distribution $\P_0^*$. We can think of $\P_0$ as marginal relative to $\P_0^*$ over $Z$ and with only $Y$ observed rather than $Y_1$ and $Y_0$.

Let $N_d$ with $d\in \{0,1\}$ denote the sample size of the trial and external datasets for $d=1$ and $d=0$ respectively. Since we are assuming $D$ is i.i.d. distributed from $\P_0$, the external sample size and RCT sample sizes are not fixed. The total sample size of the hybrid control trial $n$ is the sum of $N_0$ and $N_1$.

Let $\mu(A,X)=E[Y|X,D=1,A=a]$ for $a\in \text{support}(A)$ define the trial-specific outcome regression. Let $p(X)=P(A=1|X, D=1)$ define the trial-specific treatment propensity score --- note that the non-trial-specific treatment propensity score is always 0 since all external data consists of control units. Let $q=P(D=1)$ be the marginal probability of trial participation with $q=E[\frac{N_1}{n}]$. Let $\pi(X)=P(D=1| X)$ define the selection propensity score; this serves a similar role as the treatment propensity score but instead estimates the conditional probability of RCT trial participation given covariates $X$. Finally, let $r(X)=\frac{V[Y|X,D=1,A=0]}{V[Y|X,D=0,A=0]}$ represent the variance ratio of the outcomes between the concurrent trial controls and external controls.

\begin{table}[ht] 
\centering 
\begin{threeparttable}
\begin{tabularx}{\textwidth}{@{}lX@{}} 
\toprule Parameter & Definition \\ 
\midrule $\displaystyle \mu(A,X)$ & $\displaystyle E\bigl[Y \mid X, D=1, A=a\bigr]$, the trial‐specific outcome regression\tnote{a}. \\ 
$\displaystyle p(X)$ & $\displaystyle P\bigl(A=1 \mid X, D=1\bigr)$, the trial‐specific treatment propensity score. \\ 
$\displaystyle q$ & $\displaystyle P(D=1) \;=\; E\!\Bigl[\tfrac{N_1}{n}\Bigr]$, the marginal probability 
of trial participation. \\ $\displaystyle \pi(X)$ & $\displaystyle P\bigl(D=1 \mid X\bigr)$, the selection propensity score. \\ $\displaystyle r(X)$ & $\displaystyle \frac{V\bigl[Y \mid X,D=1,A=0\bigr]}{V\bigl[Y \mid X,D=0,A=0\bigr]}$, the conditional outcome variance ratio of concurrent 
vs.\ external controls. \\ \bottomrule 
\end{tabularx}
\begin{tablenotes}
\item[a] Under mean exchangeability discussed in the following section, $\mu(0,X) = E[Y \mid X, A=0]$.
\end{tablenotes}
\end{threeparttable}
\caption{Population parameters and their definitions.}
\end{table}

\subsection{The Trial-Specific Average Treatment Effect}

In this paper, we focus on the trial-specific average treatment effect. This measures the causal effect of the treatment $A$ on outcomes $Y$ for the population of people who were eligible to participate in the randomized controlled trial. Although other estimands of interest exist for hybrid control trials, we focus on the trial-specific average treatment effect due to its prevalence in hybrid control trial causal inference literature \cite{li_improving_2023,valancius_causal_2024, ung_combining_2024}. This parameter is also of key importance in the pharmaceutical and regulatory context where RCT trial recruitment is based on a carefully selected population, so inference would likewise need to be focused on this specific population.

% \subsubsection{The Causal Parameter}

The causal trial-specific average treatment effect is:

$$\Psi^*=E[Y_1-Y_0|D=1]$$

This estimand parallels the typical average treatment effect that would be estimated in a standard RCT. Two key assumptions are needed for identification\footnote{Some authors consider causal consistency to be an identification assumption  as well. Although we do ``assume'' causal consistency, we think of it more as a structural link between the causal and observable distributions rather than as a restriction on the set of causal data generating processes. The latter are what we prefer to call ``identification assumptions'' but this is ultimately a matter of perspective.}  of the trial-specific average treatment effect in our HCT setting:

\begin{enumerate}
    \item \textbf{Strong Ignorability}: 
    \begin{enumerate}[label=(\roman*)]
        \item{Conditional Independence}: $Y_a \perp\kern-9pt\perp A  | X, D=1$ for all $a \in A$
        \item{Positivity}: $P(A=1| X=x, D=1)>0$ for all $x \in X$
    \end{enumerate}
    \item \textbf{Mean Exchangeability of $Y_0$}: $E[Y_0|X=x, D=1]=E[Y_0|X=x, D=0]$ for all $x \in X$
\end{enumerate}

Strong ignorability is a common assumption to identify the average treatment effect when conducting inference in observational studies. That condition is met by design for the purposes of our paper since the trial is randomized. 
On the other hand, the mean exchangeability assumption is \textit{not} addressed by randomization, leaving a gap between the observable estimand that can be estimated and the causal estimand of scientific interest. This gap is precisely the bias induced by pooling in HCTs. For the time being we will proceed as if we believed this assumption. Later we will address its impact.

% \subsubsection{The Statistical Estimand}

Using the above assumptions, the following observable statistical estimand identifies the causal trial-specific ATE (see \ref{identification-proof} for proof):
\begin{align*}
    \Psi=E[E[Y|A=1, X, D=1]-E[Y|A=0, X]]
\end{align*}

An important benefit of the mean exchangeabilty assumption is that although we are interested in the trial-specific average treatment effect (thus the situation where $D=1$), the second term in the estimand above is not conditioned on $D$. Indeed, the mean exchangeability assumption ensures that $E[Y|A=0,X] = E[Y|A=0,X,D=1] = E[Y|A=0,X,D=0]$ in the context of strong ignorability. Due to this, the conditional expectation of the control subjects may be ``evaluated'' over the population of all possible HCT participants (not just the trial population). In other words, we can use the external control subjects to help estimate the trial-specific ATE even though they are not exactly from the population of interest.

\subsubsection{Efficient Estimation}

We rely on an efficient estimator $\hat{\Psi}$ for the trial-specific average treatment effect proposed by \citet{li_improving_2023}. Although others exist, we focus on this estimator because it is asymptotically efficient relative to regular and asymptotically linear alternatives. This property aligns with our goal to alleviate trial constraints since efficient estimators permit smaller sample sizes while preserving standard methods of inference and robust behavior across data-generating processes \cite{laan_targeted_2011}. This estimator is also doubly robust, making it less prone to model misspecification bias.

The efficient influence function for this estimand (as derived by \citet{li_improving_2023}) is:

 \begin{align*}
    \phi(Y,A,X,D) &= \frac{1}{q}\left[D(\mu(1,X)-\mu(0,X)-\Psi) +\left(\frac{DA}{{p}(X)}-{W}(A,X,D)\right)(Y-\mu(A,X))\right]\\
    {W}(A,X,D) &=\frac{D(1-A)+(1-D){r}(X)}{{\pi}(X)(1-{p}(X))+(1-{\pi}(X)){r}(X)} \pi(X)
\end{align*}

The derivation of this influence function takes into account that $P(A=0|D=0) = 1$, i.e. that the external data contains only controls. Explicitly incorporating this knowledge actually increases the efficiency of the analysis (i.e. is expected to give smaller confidence intervals and p-values).

Solving an empirical score equation based on this influence function yields the following efficient estimator for the trial-specific average treatment effect (with $E_n$ estimating the population expectation with a sample average): 

\begin{align*}
    \hat{\Psi}&= \frac{1}{\hat{q}}E_n\left[D(\hat{\mu}(1,X)-\hat{\mu}(0,X)) +\left(\frac{DA}{\hat{p}(X)}-\hat{W}(A,X,D)\right)(Y-\hat{\mu}(A,X))\right]\\
    \hat{W}(A,X,D) &=\frac{D(1-A)+(1-D)\hat{r}(X)}{\hat{\pi}(X)(1-\hat{p}(X))+(1-\hat{\pi}(X))\hat{r}(X)} \hat \pi(X)
\end{align*}
In our setting, this is the equivalent of the popular augmented inverse propensity weighted (AIPW) estimator that is used in observational studies. It is also (when calculated with sample-splitting) the double machine learning (DML) or ``efficient estimating equations'' estimator for our setting \cite{chernozhukov2024doubledebiasedmachinelearningtreatment}.
Although we use DML here for accessibility and consistency with \citet{chernozhukov_long_2024}, the work most closely related to ours, other (asymptotically equivalent) approaches for efficient estimation such as targeted maximum likelihood estimation (TMLE) can also be used \cite{laan_targeted_2011}.

Given that we are selecting $q$ by design of the HCT experiment, the estimation of $\hat{\Psi}$ depends primarily on the estimation of functional components $\mu(A,X)$, $p(X)$, and $\pi(X)$ as well as the variance ratio $r(X)$. In the formula for the efficient estimator, $\hat \mu(A,X)$, $\hat p(X)$, $\hat \pi(X)$, and $\hat r(X)$ are machine learning estimates of their corresponding population objects, converging at assumed $L^2$ rates. 

Under general conditions, $\hat\Psi$ is asymptotically normal with large-sample variance of $\sqrt{n}\hat\Psi$ given by $E[\phi(O)^2]$, which can be approximated with a plug-in estimator using the appropriate estimated regressions. This facilitates the construction of confidence intervals and p-values.

When one assumes mean exchangeability and has a binarized outcome $Y$, $r(X)$ reduces to 1. In this paper, we will restrict the outcome to be binary so that $r(X)$ does not require estimation and the results can be presented more economically. However, the results should still extend well to other cases (e.g. continuous or even censored outcomes). For example, the value of $r(X)$ only affects efficiency and not unbiased estimation, and $r(X)$ can be estimated well in a scenario where a continuous outcome is of interest as shown in \citet{li_improving_2023}.

\section{Methods}
\label{sec3}
The issue at hand is that this estimator, like others, can only target the observable estimand which relies on the strict mean exchangeability assumption in order to get any benefit from the external control subjects. Although there are approaches such as the test-then-pool method to detect violations of this assumption, they cannot recognize every violation \cite{viele_use_2014}. Further, while preliminary tests may identify a violation of the mean exchangeability assumption, it's possible that the violations only lead to small amounts of bias. In these cases, the external data may have been useful, providing greater benefits than costs, but not used due to the black-and-white nature of significance tests.  More complicated methods of dynamic borrowing include Bayesian methods, but simulation studies have shown that power gains are limited when strict type-one error rates are enforced \cite{koppschneider_power_2020}. Due to the inevitability and uncertain quantification of this bias, hybrid control trials are typically not accepted by major regulatory bodies \cite{noauthor_international_2000}. 

Considering these issues, we shift focus away from\textit{ if }bias is present to \textit{how much} bias is present when this assumption is violated. We argue that the benefits of data borrowing in HCTs can lead to overall increased trial feasibility even if the external data are slightly biased away from the trial data. Therefore, there is great interest in quantifying this bias and bounding type-one error inflation so that researchers can safely reap the benefits of hybrid control trials, including smaller sample sizes, shorter timelines, and reduced budgets. This motivates the following analyses.

\subsection{Omitted Variable Bias}

Any violation of the mean exchangeability assumption can be seen as a direct result of omitted variable bias. In the instance that $E[Y_0|X,D=1]\neq E[Y_0|X,D=0]$, we should imagine that mechanistically there is some unobserved confounder $Z$ that induces the difference in the outcome's dependence on the observed covariates. 

As an extreme and unrealistic example, consider a hypothetical hybrid control trial where a randomized controlled trial is conducted today and the external controls are collected from a previous trial. Suppose that this external data was collected far in the past, such as a century ago. Even if they share information on covariates $X$ such as age, gender, weight, etc., it is no surprise if the average outcomes in a particular covariate stratum differ between the trial controls and external controls, i.e. $E[Y_0|X, D=1]\neq E[Y_0|X, D=0]$. One can deduce that there may be confounding simply due to the stark differences between the modern time and the time when the external data were collected. However, as relevant information that captures these time- and place-dependent differences is collected --- such as sanitary conditions, occupations, quality of diet, exposure to pollutants, among many others --- the conditional expected outcomes should begin to converge. If all relevant confounders, consisting of the sets $X$ and $Z$ (originally unmeasured), are somehow able to be measured, the conditional expected outcomes will become equivalent between our control groups. We can call this full set of confounders $\tilde{X}$. 

This idea generalizes to any hybrid control trial and recontextualizes the issue to be an omitted variable bias problem. In particular, $\Psi$ no longer identifies $\Psi^*$ since it does not include the relevant covariates. Instead, $\tilde{\Psi}$, denoting the statistical estimand that conditions on $\tilde{X}$, identifies the true causal parameter $\Psi^*$. This discrepancy introduces the bias $b = |\Psi - \Psi^*|=|\Psi - \tilde\Psi|$. By bounding this bias, we can estimate how far $\Psi$ is from $\tilde{\Psi}$ when unobserved confounding is present. In the rest of this paper, we also use the tilde to denote long regressions (e.g. $\tilde\mu(A,\tilde X) = E[Y|\tilde X, A=a, D=1]$) in which the full set of confounders $\tilde X$ is used instead of the short regressions in which only the observed covariates $X$ are used.

In what follows, our goal will be to find an upper-bound $|\Psi - \Psi^*| \le B$, quantifying the maximum bias that could be introduced if the mean exchangeability assumption were violated. We will then derive an estimator $\hat B$ for $B$, the bias bound. 

\subsection{The Bias Bound $B$}

We leverage prior omitted variable bias methodologies for developing sensitivity analyses based on the Riesz representation form of the causal estimand \cite{chernozhukov_long_2024}. Prior sensitivity analyses have been developed for a variety of estimands in observational studies such as the average treatment effect and weighted average incremental effects. In particular, this method works for a class estimands that can be written as a linear functional of the conditional expectation function of the outcome, also known as the outcome regression. These estimands can be written in the form: $\Psi(P_0)=E[m(\mu)(O)]$ where $\mu$ is the outcome regression based on $P_0$ and $m$ is some linear functional. 

In the case of the trial-specific ATE, we can use conditioning arguments to write the estimand as $E[m(\mu)(O)]$ where $m(\mu)$ returns the function $O \mapsto q^{-1}D(\mu(1,X)-\mu(0,X))$ (see \ref{rep} for proof).

\subsubsection{Riesz Representation}

The Riesz representation theorem states that within a Hilbert space any continuous linear function that maps to a real number can be written as the inner product of the argument to the function and some other element in the Hilbert space \cite{intro, lee2025rieszboostgradientboostingriesz}. Using this theorem, we can then rewrite estimands of the form given above as $E[m(\mu)(O)]=E[\mu(O)\alpha(O)]$ where $\alpha$ is defined as the Riesz representer. 

As an example, we can consider the well-known estimand for the average treatment effect of a binary treatment: $\Psi(P_0)=E[E[Y|A=1,X]-E[Y|A=0,X]]$. The ATE can be defined as a linear functional of the outcome regression with $m(\mu)(O)= \mu(1,X)-\mu(0,X)$. Standard conditioning arguments show that the Riesz representer for the ATE is $\frac{A}{P(A=1|X)}-\frac{1-A}{P(A=0|X)}$, the usual inverse propensity weights. In this way, the Riesz representation motivates the use of such weighted estimators \cite{lee2025rieszboostgradientboostingriesz}. 

\citet{chernozhukov_long_2024} uses this representation to define a bias bound $B$ for any Riesz-representable estimand, dependent on the residuals between the long and short formulations of the outcome regression and Riesz representer. Since the short regressions are projections of the long regressions onto a smaller space of functions depending only on $X$, these short regressions are orthogonal to the additional regression components living outside of this smaller subspace. Relying on the orthogonality of $\mu$ and $\alpha$ on these residuals, it is shown that the bias (in terms of the difference of the long and short estimands) is bounded: $\Psi - \tilde\Psi=E[(\mu(O)-\tilde \mu(O))(\alpha(O)-\tilde \alpha(O))] \le \|\mu - \tilde\mu\|\| \alpha-\tilde\alpha\|$. That is: the maximum bias is the covariance of the approximation errors of the short representations, which is upper bounded by the product of the $L^2$ errors. The implication of this result is that there is a sharp bias bound that doesn't depend on the full complexity of the (``long'') data-generating distribution, only on how well the short regressions approximate the long regressions in mean-squared error. Another nice implication of this result is that we can concern ourselves only with confounders that affect both the outcome regression $\mu$ and the Riesz Representer $\alpha$, given the product term in the expectation. If a confounder only affects one of the terms, say $\mu$, the bias will be equal to zero since $\alpha(O)-\tilde \alpha(O)$ will also be equal to zero. This implication will later be further discussed in the context of HCTs.

We are able to extend these findings to the hybrid control trial scenario in which we use the trial-specific average treatment effect, which is a partially linear functional of the conditional expectation function of the outcome. The following fits the trial-specific ATE into the methodology proposed by \citet{chernozhukov_long_2024}. 

\subsubsection{Deriving the Riesz Representer}

First, we identify the Riesz representer using the Riesz representation theorem, showing that $\Psi=E[E[Y|A=1, X, D=1]-E[Y|A=0, X]]$ is equal to the inner product of the input outcome regression $\mu(A,X) = E[Y|A=a,X,D=1]$ (also equivalent to $E[Y|A=a,X,D=0]$ by the mean exchangeability identification assumption when A=0) and some Reisz Representer $\alpha(A,X,D)$.
In \ref{rrproof}, it is shown that the Riesz representer of the trial-specific average treatment effect is
$$\alpha(A, X,D)=\frac{1}{q}\left(\frac{DA}{p(X)}-\frac{D(1-A)+(1-D){r}(X)}{{\pi}(X)(1-{p}(X))+(1-{\pi}(X)){r}(X)} \pi(X) \right).$$ 

\subsubsection{Measuring Unobserved Confounding}
\label{sec3.2.3}

The methodology proposed by \citet{chernozhukov_long_2024} relies on the outcome regression and the Riesz representer, which have been identified above, and gains in explanatory power from latent, or unobserved, confounders. We can formally define $C^2_Y$ and $C^2_D$ as the gain in information from including unobserved confounders $Z$ in the outcome regression and Riesz representer in terms of the non-parametric $R^2$. We let the tilde denote the long regression in which $\tilde X$, the set of observed covariates and unobserved confounders, is used in the estimation of the functional components. The functional components of the Riesz Representer in this case are $r(X)$, $p(X)$, and $\pi(X)$. Since $r(X)$ reduces to 1 and $p(X)$ is roughly set to some value in an RCT, the primary gains in explanatory power in the Riesz Representer likely come from estimation of $\pi(X)$. We can thus think of the gains in explanatory power of the Riesz Representer primarily focusing on the gains in explanatory power a confounder may have on the trial selection propensity $\pi(X)$ -- note that this is not fully the case but we believe that this may be helpful for understanding what $C^2_D$ measures. In other words, $C^2_D$ roughly measures how well the observed covariates capture the differences between the external and RCT populations.

\begin{align*}
    C^2_D&= \frac{E[\tilde \alpha(A,\tilde X,D)^2]-E[\alpha(A,X,D)^2]}{E[\alpha(A,X,D)^2]}\\
    C^2_Y&= \frac{E[(\tilde \mu(A,\tilde X)-\mu(A,X))^2]}{E[(Y-\mu(A,X))^2]}
\end{align*}

Note that the non-parametric partial $R^2$ is not bounded similarly to a typical measure of $R^2$. The gains in explanatory power are relative to what is already explained by the set of observed covariates $X$; if the gains from including confounders exceed what is already explained, the numerator of the equations for $C^2_Y$ and $C^2_D$ will be larger than the denominators. Therefore, it is possible for these values to exceed 1 unlike a typical measure of $R^2$. Further, since the measure is relative to what is already captured by the observed covariates $X$, it is possible that some confounder $Z$ has a strong relationship with the outcome or Riesz Representer but a small non-parametric $R^2$ if a covariate exists in $X$ that already captures some mechanism of $Z$ (ex. a confounder such as diet may have a strong relationship with some outcome but part of the relationship may be captured by an observed covariate such as BMI or cholesterol levels).

\subsubsection{Calculating the Bias Bound $B$}

As in \citet{chernozhukov_long_2024}, we reformulate the squared-bias as

\begin{align*}
    |\Psi-\tilde \Psi|^2 &=\rho^2 C^2_YC^2_DS^2\\\
    S&= \sqrt{E[Y-\mu(A, X)]^2E[\alpha( A, X, D)]^2}\\
    \rho&=Cor(\mu-\tilde \mu, \alpha-\tilde \alpha)
\end{align*}
where $\rho$ measures the degree to which confounding in $\mu$ is related to confounding in $\alpha$.

The quantity $S$ depends only the observable (``short'') distribution, whereas the other parameters depend on the long distribution. Therefore only $S$ can be estimated from the data. $\rho$, $C^2_Y$, and $C^2_D$ must be assumed or bounded on the basis of domain knowledge, thus we will refer to these as the sensitivity parameters. There are, however, sensible ways of setting these parameters. It is possible to pick a good value for $\rho$ --- as well as $C^2_D$ and $C^2_Y$ --- empirically through benchmarking methods using the observed covariates. For example, given a dataset with multiple observed covariates, one can select an observed covariate to be treated as unobserved and calculate estimates for $m$, $\tilde m$, $\alpha$, and $\tilde \alpha$ to then use to estimate $\rho$, $C^2_D$, and $C^2_Y$ through a ``leave-one-out" approach \cite{mcclean_calibrated_2024}. The decision of which variable to treat as unobserved depends on multiple factors such as the nature of confounding we might expect to be present as well as the potential magnitude of confounding. However, there is no free lunch: any method for choosing these parameters incurs some risk and therefore requires a good understanding of the problem domain. In practice these parameters should also be varied over plausible ranges --- see Section \ref{sec3.4} on critical causal gaps below.

Finally, also note that the bias bound relies on products of the sensitivity parameters. Therefore, if any of these are $0$, the resulting bias from confounding is also $0$. Similarly, if any one of the parameters is large but the others are `small', the resulting bias most likely will also be relatively small. In the context of HCTs, this translates to confounding must jointly affect (measured by $\rho$) both the outcome regression and Riesz Representer. Therefore, confounders must affect both trial participation and the outcome. Practically speaking, if external data is chosen well, we can somewhat control the magnitude of $C^2_D$ and $B$. In theory, even if there exists a confounder which greatly influences the outcome, if the confounder does not explain differences in the trial and external control populations, bias will not be incurred. Similarly, if there exists a confounder which greatly influences the Riesz Representer/trial participation but not the outcome, bias will not be incurred -- though this is less controllable through selection of external controls, but may be useful in the discussion of possible confounders.

In our paper, we proceed with the most extreme estimate of $|\rho|=1$ for a conservative estimate of $B$. As a result, our bias bound is
$$B=C_YC_DS.$$
Under adversarial confounding when $\rho^2=1$, the bias bound is sharp with $B=C_Y C_D S$. 
In our simulations, we proceed with the assumption that $C_D$ and $C_Y$ have been well-chosen a-priori (e.g. using one of the methods described above) since we want to separately evaluate the performance of our method from a user's knowledge or assumptions. Certainly, our method will appear to fail if these parameters are poorly chosen.

\subsubsection{Estimator of the bias bound}

In order to estimate $B$, we must first estimate $S$, which depends on our Riesz representer and outcome regressions. For the outcome regression $\mu(A,X)$, we estimate separate regressions for each treatment. To estimate $\mu(1,X)$, we use only the data corresponding to those who received treatment, regressing $Y$ onto $X$ for only $A=1$ (note this implied $D=1$). However, under the mean exchangeability assumption, we estimate $\mu(0,X)$ using both the RCT and external control arms (this is where we gain efficiency in the estimation of the trial-specific ATE). This corresponds to regressing $Y$ on $X$ for $A=0$, regardless of the value of $D$. Although such pooling introduces bias when mean exchangeability fails (since $\mu(0,X)$ is no longer identified by $E[Y|A=0,X]$), this is precisely the bias that we are measuring in this paper. If we can discern the magnitude of the bias, we can evaluate whether the potential bias from pooling outweighs the efficiency gains it provides.

To estimate $\alpha(A,X,D)$, we use a plug-in estimator based on the functional components outlined in Section \ref{sec2}. We first use machine learning to estimate $\pi(X)$ and $p(X)$ by respectively regressing $D$ onto $X$ for all participants and $A$ onto $X$ only for those who participated in the RCT. We can then use these to estimate $\alpha(A,X,D)$ for every participant in the HCT, plugging in values into the equation for $\alpha$.

Each of the regression tasks we describe can be accomplished with generic nonparametric or machine learning methods, e.g. random forests, boosting, neural networks, kernel ridge regression, elastic net, etc. When we have estimated $\alpha$ and $\mu$, we can then plug these in to estimate $S$ such that $\hat{S}= \sqrt{E_n[Y-\hat\mu(A,X)]^2E_n[\hat{\alpha}(X, A, D)]^2}$. This can then be used in the estimation of the bias bound, plugging in $\hat{S}$ with plausible hypothetical values for $C_Y$ and $C_D$.

\subsubsection{Inference on the bias bound}

Since the bias bound requires estimation of the term $S$, it is itself a random quantity with some uncertainty that should be quantified. Inference on $B$ (via $S$) can be conducted using the delta method to combine estimates of the standard errors of $\nu^2 = E[\alpha(X,A,D)]^2$ and $ \sigma^2=E[Y- \mu(A,X)]^2$. The influence functions for $\nu^2$ and $\sigma^2$ are $\phi_{\sigma^2}=(Y-\mu(A,X))^2-\sigma^2$ and $\phi_{\nu^2}=\alpha(X,A,D)^2-\nu^2$. See \citet{chernozhukov_long_2024} for additional details. The influence function of $B$ is then:

$$ \varphi_B=\frac{|\rho|}{2}\frac{C_Y C_D}{S}(\sigma^2\phi_{\nu^2}+\nu^2\phi_{\sigma^2}).$$
Given the influence function, the variance of $\sqrt{n}B$ is $E[\varphi_B(O)^2]$.  We estimate variance with a simple plug-in estimator, using estimated versions of each of the required population quantities (e.g. $\hat\psi_{\nu^2}=\hat\alpha(X,A,D)^2-\hat\nu^2$).

\subsection{Using the Bias Bound}

Presuming that we had a true upper bound $B \ge b$ and the true value of the observable trial-specific ATE $\Psi$, we could define a range $[\Psi_{-}, \Psi_{+}]$ (calculated using $\Psi_\pm=\Psi \pm B$) which would be certain to include the true, causal, trial-specific ATE. Our task is therefore to produce estimates and conduct inference for the estimands $\Psi_\pm$ using our efficient estimates of $\Psi$ and $B$ which are described above.

\subsubsection{Inference on $[\Psi_-, \Psi_+]$}

Since the target has changed from estimating $\Psi$ to $\Psi_\pm$, inference likewise changes. The influence functions of the bounds $\Psi_\pm$ can be derived using the delta method, relying on the influence functions of $\Psi$ and $B$, as follows: 
$$\varphi_\pm=\phi \pm \varphi_B $$
where $\phi$ is the influence function for $\Psi$ defined in section \ref{sec2}.  The variance of the upper and lower bounds, $\sigma^2_+$ and $\sigma^2_-$, can thus be found taking the variance of $\varphi_\pm$.

Instead of constructing a two-sided confidence interval around the estimates of the bounds (such as one would do for inference on $\hat{\Psi}$), we follow \citet{imbens} to construct one-sided confidence bounds on $\Psi_+$ and $\Psi_-$ that maintain uniform asymptotic coverage across $B$. 

\begin{align*}
        &CI_{\eta}=\left[ \hat{\Psi}_{-} -\frac{c_{\eta}\hat{\sigma}_-}{\sqrt{n}},\hat{\Psi}_{+} +\frac{c_{\eta}\hat{\sigma}_+}{\sqrt{n}}\right]\\
     &\text{with $c_{\eta}$ such that}\\
    &\Phi\left( c_{\eta}+\frac{\sqrt{n}(\hat{\Psi}_{+}-\hat{\Psi}_{-})}{\max\{\hat{\sigma}_-, \hat{\sigma}_+\}} \right)-\Phi(-c_{\eta})=1-\eta\\
    &\text{where $\Phi(c)=\int_{-\infty}^c\frac{1}{2\pi}e^{\frac{-x^2}{2}}dx$.}
\end{align*}

Given a critical value of $\eta$,  this confidence region that uses both confidence bounds guarantees coverage of $\Psi$ to asymptotically approach $1-\eta$ uniformly in $B$. This is important when there is very little unmeasured confounding. Another benefit of this procedure is that whenever $B\approx0$, the confidence region becomes equivalent to that of a naive HCT analysis with a standard two-sided pointwise confidence interval. More details and justification can be found in \citet{imbens}.

\subsection{Identifying the Critical Causal Gaps}
\label{sec3.4}

In reality, the true values of $C^2_Y$ and $C^2_D$ will always be unknown since they depend on the unobserved confounders $Z$. Based on different combinations of $C^2_Y$ and $C^2_D$, it is possible to identify values of $C^2_Y$ and $C^2_D$ where any significant results from the HCT analysis are invalidated by the potential magnitude of bias present. These are values that result in the confidence bounds of $\hat{\Psi}_{+,-}$ crossing the null value (in many cases this is 0, indicating that there is no trial-specific average treatment effect). A robustness value can also be calculated as a summary of robustness against bias \cite{chernozhukov_long_2024}. The robustness value measures the minimum common strength of confounding needed (where $C^2_D = C^2_Y$) to invalidate significant findings from the HCT, given some $\rho$. For example, if the robustness value is 0.1, we know that any combinations of $C^2_Y$ and $C^2_D$ where both are greater than 0.1 will result in insignificant findings. Thus, when the robustness value is relatively low, it indicates there is a large risk of bias from using the external control data. To contextualize if the robustness value is relatively `low' or `high', observed covariates can be used to obtained benchmarked values for comparison. Importantly, the robustness value is a sufficient condition, not a necessary condition, for statistical insignificance. Combinations where either $C^2_D$ or $C^2_Y$ fall beneath the robustness value may still invalidate significant findings if the other parameter is sufficiently large, due to the calculation of the bias bound involving the product $C_YC_D$.

We can again extend the work of \citet{chernozhukov_long_2024} to create contour plots to display these combinations of $C^2_Y$ and $C^2_D$, such as the following:

\begin{figure}[H]
    \centering
    \includegraphics[width=0.9\textwidth]{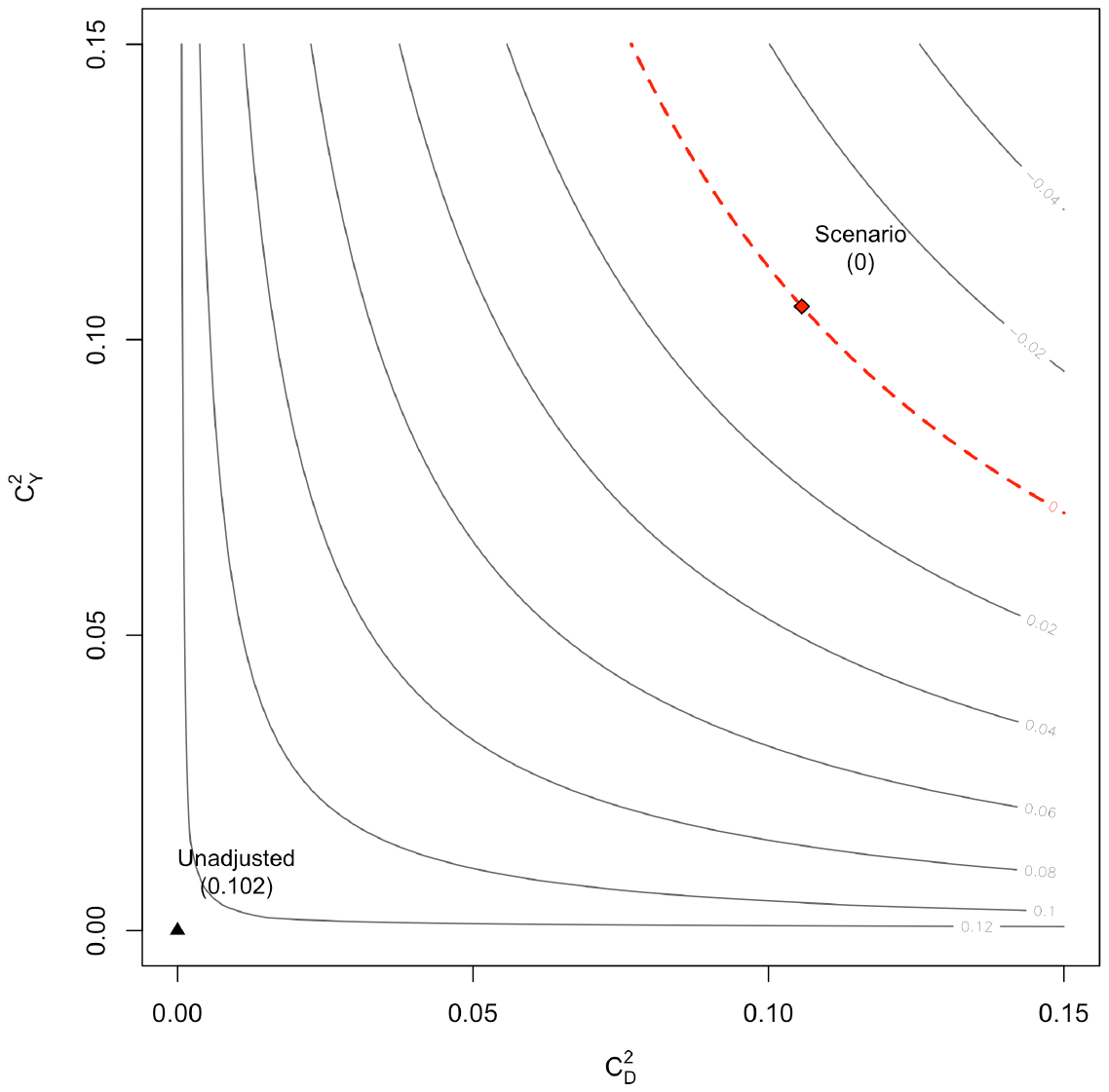}
    \caption{A contour plot demonstrating robustness of a hypothetical HCT. The x-axis shows increases in explanatory power from Z in the Riesz Representer, and the y-axis shows increases in explanatory power from Z in the outcome regression. The unadjusted (naive) HCT analysis lies at the origin and the dotted red line establishes the threshold where 
 $B$ negates significant findings. Different combinations in $C^2_D$ and $C^2_Y$ can also be specified, as indicated by the red diamond. Here, we use the robustness values as the scenario, demonstrating one point at which results become insignificant. The lower confidence bound of the trial-specific ATE is shown in parentheses.}
    \label{fig:contour}
\end{figure}

Given a similar plot, researchers should be able to identify plausible increases in $R^2$ for the outcome regression and the Riesz representer through professional knowledge of the topic or other methodologies for omitted variable bias. Some of these methodologies include algorithms such as the ``leave-one-out" approach introduced earlier and others explored in section \ref{sec5}. Conclusions can then be made based on the robustness of evidence from the HCT at the researcher's discretion.  Although it may be impractical to assume that researchers can pinpoint the exact amount of unobserved confounding, it is plausible that researchers can assess robustness against bias due to the flexible nature of the contour plots. If the data are shown to not be robust against bias, it may be best to proceed with a pure RCT analysis. Particularly, if the plausible increases in $R^2$ far exceed the critical causal gap values, the risk of introducing bias exceeds the potential benefits of incorporating external data.

\subsection{Summary}

In summary, the steps one would take to execute this analysis are as follows: 

\begin{enumerate}
    \item Gather external control data with measurements on the same outcome variable and covariate variables as the RCT data. Through some preliminary analysis (examples in Section \ref{sec5}), check that the covariate distribution of the two sets of control data are not significantly dissimilar.
    \item Use double debiased machine learning to estimate $S$ and $\Psi$ as described above.
    \item Pick $\rho=1$ (for conservative estimates) and for plausible values or ranges of $C^2_Y$ and $C^2_D$:
    \begin{enumerate}
        \item Calculate the bounds $\Psi_\pm$ using the estimates of $\Psi$, $S$, and posited values of $C^2_Y$ and $C^2_D$.
        \item Estimate one-sided confidence bounds using the adjustment for when $B\approx0$ and standard error of the bounds found by combining the variances of $\Psi$ and $B$.
    \end{enumerate}
    \item Optional: Create contour plots to visualize results and/or calculate robustness values for flexible interpretation.
\end{enumerate}

\section{Simulation Study}
\label{sec4}
In the following section, we use simulations to verify the performance of the sensitivity analysis as well as to explore the practicality of aiming for improved efficiency even with the wider confidence bounds resulting from the sensitivity analysis. In particular, it is shown that the true bias is correctly and consistently bounded by the bias bound $B$ in varying scenarios, type-one error inflation is corrected by the sensitivity analysis, and power can be improved even after accounting for potential omitted variable biases. 

\subsection{Data Generating Process}

The following data generating process (DGP) was used:

\begin{align*}
    D &\sim \text{ Bern}(p=q)\\ 
    X &\sim \text{ Bern}(p=0.6)\\
    Z|D &\sim \begin{cases}
        \text{ Bern}(p=0.3) \text{ when } D =1\\
        \text{ Bern}(p=\zeta) \text{ when } D=0\\    
    \end{cases}\\
    A|D &\sim \begin{cases}
        \text{ Bern}(p=0.5) \text{ when } D=1\\
        \text{ } 0 \text{ when } D=0\\
    \end{cases}\\
    Y^*|A, X, Z &=9.5+\beta A-0.45X+0.75Z+\mathcal{N}(0,1)\\
    Y &= \begin{cases}
        1 \text{ when } Y^*\ge10\\
        0 \text{ when } Y^*<10
    \end{cases}
\end{align*}

 Note that the DGP is not meant to be realistic, but rather be relatively simple so that the operating characteristics of the sensitivity can be understood under easily-manipulated scenarios. In this scenario, Y is a binarized outcome, indicating that some measurement of interest $Y^*$ was greater than or equal to a cutoff value of 10. A binary outcome was chosen to proceed with the simplified formulation of the Riesz representer $\alpha$. Several values in the simulation are variable: $q, \zeta, \beta $, allowing manipulation for different scenarios. As defined earlier, $q$ is the probability of trial participation, which equates to $E[\frac{N_1}{n}]$. Throughout all simulations, $q$ is varied from 0.1 to 0.9 with step size 0.1. $\zeta$ is the probability of some binary unobserved confounder Z for external control participants. Bias can be manipulated by leveraging the distribution of the unobserved confounder $Z$ and the size of the external control arm $N_0$. $\beta$ is the effect of treatment on the measurement of $Y^*$. By changing this value, we are able to assess scenarios in which there is a true treatment effect and in which there is a true null treatment effect.

\subsection{Simulation Setup}

For the following simulations, R 4.4.2 was used with seed 509.

We compare 3 algorithms: RCT, HCT (we also refer to this as a `Naive' HCT analysis), and HCT + sens (HCT supplemented with sensitivity analysis). For the RCT analysis, we use the same generated data as the HCT analysis but only use those who are trial participants ($D=1$). Instead of calculating the trial-specific ATE estimator, an efficient estimator for the ATE is used: $\Psi_{ate}=E \left[ \mu(1,X)-\mu(0,X)+\left(\frac{A}{p(X)} - \frac{1-A}{1-p(X)}\right)(Y-\mu(A,X)) \right]$. In the RCT analysis, $\mu(0,X)$ is estimated separately from the HCT scenario outcome regressions, using only the trial control data. $95\%$ point-wise confidence intervals are calculated for $\hat{\Psi}_{ate}$. For the HCT analysis, we calculate the trial-specific ATE using the estimator $\hat{\Psi}$. $95 \%$ point-wise confidence intervals are also calculated on $\hat \Psi$. For the HCT + sens analysis, we not only calculate the trial-specific average treatment effect but also the bias bound $B$ to estimate $\Psi_\pm$. In the estimation of the bias bound $B$, the parameter $\rho$ was set to 1 for conservative estimates. Confidence bounds were calculated to create $95\%$ confidence regions for the HCT + sens analysis using estimated values of $\Psi$ and $B$ as outlined in Section \ref{sec3}.

 For all algorithms, we rely on double debiased machine learning using the ranger algorithm \cite{chernozhukov2024doubledebiasedmachinelearningtreatment}. Double debiased machine learning was used for the estimation of $\Psi$, $\Psi_{ate}$, and $S$ (the component in the bias bound not dependent on the parameters $\rho$, $C^2_D$, and $C^2_Y$). Random forests were chosen to estimate the functional components of these statistical estimands due to their ability to perform quickly and well without much tuning.

A key assumption used in the simulations is that the true values of $C^2_Y$ and $C^2_D$ are known. Using the DGP, we can closely estimate these values by generating hypothetical trials with 1,000,000 participants. The non-parametric $R^2$ can then be compared between scenarios in which only $X$ is used for estimation and in which the full set of relevant confounders $\tilde X$ are used to find the true values of $C^2_Y$ and $C^2_D$ using the formulation in Section \ref{sec3}. While this is not possible in real HCT trials, the aim of the simulations is to show that the sensitivity analysis is reliable, given we have some sufficient understanding of the covariate-outcome relationship.

While the simulations that follow use specific values of RCT and external arm sample sizes, the sample sizes of these groups in each iteration are ultimately determined by the randomness of the DGP and $q$. For example, if we want to explore a scenario in which 100 RCT participants are desired, the total HCT sample size $n$ is first found by dividing 100 by $q$. Then the DGP randomly assigns the $n$ participants to either the RCT or external arm. Thus, the RCT sample size fluctuates loosely around the desired sample size of 100. Note that the rest of the paper refers to these desired RCT and external sample sizes with the understanding that each iteration will roughly fall near these desired sample sizes since these are not actually fixed in the DGP.

\subsection{Verifying the Bias Bound}
\label{sec:biasbound}

\begin{figure}[h]
    \centering
    \includegraphics[width=1\textwidth]{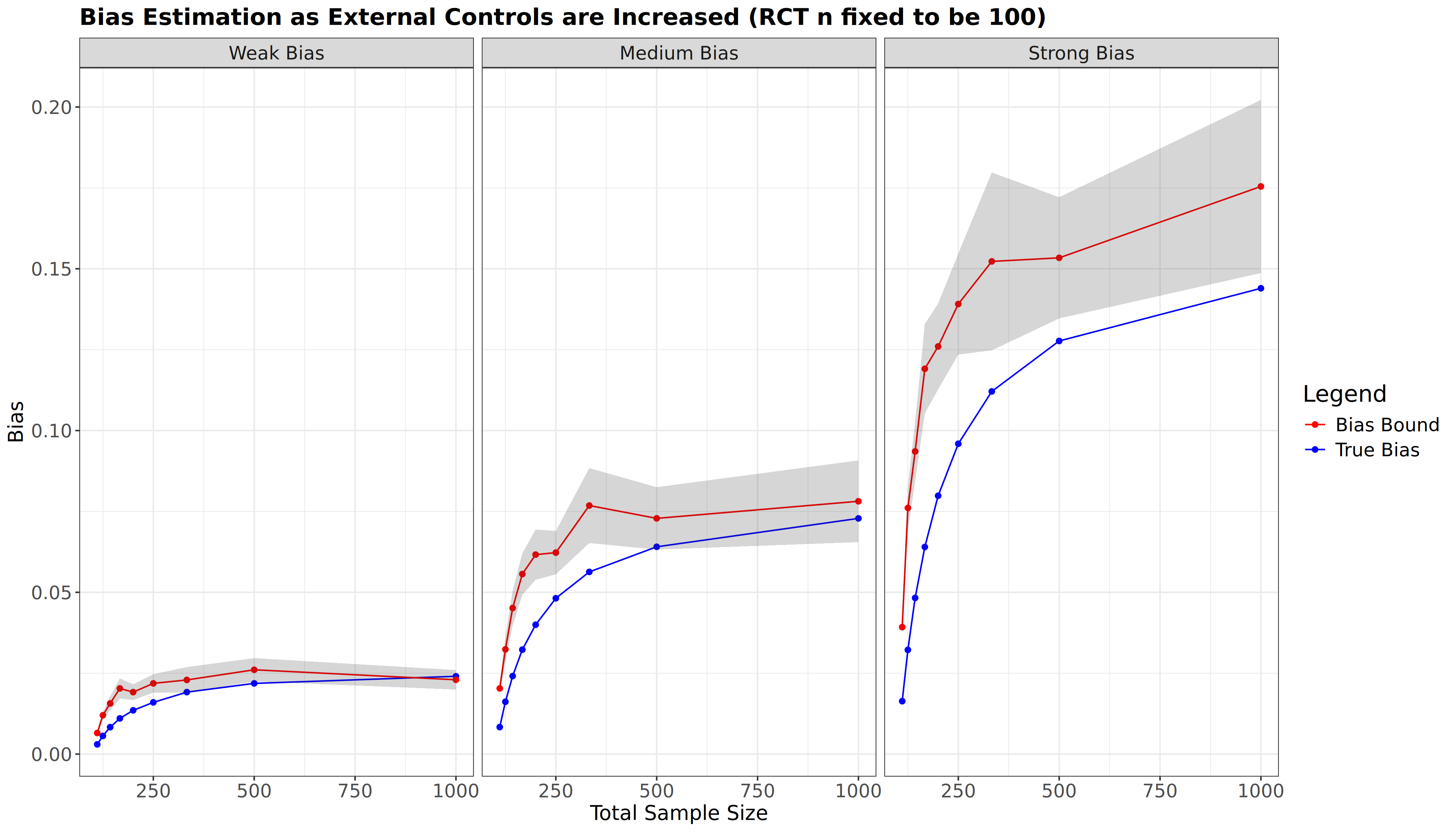}
    \caption{True bias $b$ (shown in blue) compared against the calculated bias bound $B$ (shown in red). Grey areas represent the $95 \%$ confidence intervals for the estimated bias bound. Total HCT sample size is varied with the RCT sample size $N_1$ fixed to be roughly 100. The leftmost panel shows weak bias in Z, middle shows medium bias in Z, and rightmost panel shows strong bias in Z.}
    \label{fig:bias}
\end{figure}

In Figure \ref{fig:bias}, the terms ``strong bias", ``medium bias", and ``weak bias" are used. These indicate how different the distribution of the unmeasured confounder $Z$ is between the trial participants and the external control participants. The RCT sample size is set to roughly 100 with varying amounts of external data, determined by $q$. The total sample size can then be calculated by dividing the input RCT sample size by each value of $q$. The sensitivity analysis was applied to the data to obtain estimates of $B$ and their corresponding 95\% confidence intervals for each simulated scenario. Recall from earlier that the magnitude of the true bias $b=|\Psi - \Psi^*|= |E\bigl[(E[Y |A = 1, X, D = 1] - E[Y |A = 0, X]) - E[Y_1-Y_0|D=1]\bigr] |$. 10,000,000 hypothetical participants were simulated using the DGP to approximately calculate the true value of $b$.

\begin{table}[h]
\centering
\begin{tabular}{ |p{3cm}||p{3cm}|p{3cm}|p{3cm}|  }
 \hline
 \multicolumn{4}{|c|}{Scenarios:} \\
 \hline
 Variable&Weak Bias&Medium Bias&Strong Bias\\
 \hline
 $\zeta$   & 0.4    &0.6&   0.9\\
 $\beta$&   0.75  & 0.75   &0.75\\
 \hline
\end{tabular}
\caption{Values used in the simulation for Figure \ref{fig:bias}. The strength of bias is varied by manipulating the underlying distribution of $Z$ in the external control group in comparison with the underlying distribution of $Z$ in the RCT group}
\end{table}

\subsubsection{Bias can be correctly estimated with any amount of external data but increases as external data size grows}

The estimation of the bias bound $B$ in the sensitivity analysis consistently correctly bounds the true bias $b$ with $B>b$ in almost every simulated scenario, regardless of the ratio of external data to trial data and the strength of bias arising from $Z$. The one exception is under weak bias and $n=1000$, but the bias bound estimate is only slightly lower than the true bias with the confidence bounds containing the true value.

As expected, the size of the true bias $b$ increases both as the difference in the distributions of $Z$ increases and the size of the external control arm increases. The increase in bias as the external control arm increases can be explained by the increase in amount of biased data, increasing the influence of the bias. The bias bound $B$ follows a similar shape in response to the trajectories of $b$. Interestingly, the standard error of the bias bound $B$ also increases as the size of the external control arm increases. Since the standard error of the bias bound is partially a function of the expectation of $\alpha$, it is driven by the $\frac{1}{q}$ term, inflating as $q$ becomes closer to 0 (equivalently as $n_0$ increases relative to $n_1$). Given this, one should be hesitant on using an overwhelming amount of external controls in comparison to the RCT sample size and even more so if the true effect size is expected to be relatively small.

\subsection{Type I Error Control}

In the second figure, type I error rates are calculated. Given the interest in type I error rates, $\beta$ is set to 0, indicating a null treatment effect. Two scenarios of ``No Bias'' and ``Bias'' are explored by manipulating $\zeta$. The simulation of ``Bias'' here is meant to reflect a scenario where there are only small deviations in the distribution of $Z$ because larger deviations can be identified in preliminary assessments, indicating that the external data are not fit to be used with the RCT data. External sample sizes of roughly 100, 500, and 1000 are tested across varied RCT sample sizes (ranging from 50 to 250 by increments of 50), altering the value of $q$ accordingly. Three analysis methods are compared: pure RCT analysis (with no external controls), naive HCT analysis, and HCT analysis combined with the sensitivity analysis (HCT + sens.). Type I errors were determined using the $95\%$ confidence regions for the HCT + sens. method and $95\%$ confidence intervals for the pure RCT and naive HCT analysis. Simulations with 1000 iterations were used to estimate the type-one error rates. $95\%$ confidence intervals are also calculated for the estimates of type I error, using the variance of binary outcomes.

\begin{table}[h]
\centering
\begin{tabular}{ |p{3cm}||p{3cm}|p{3cm}|  }
 \hline
 \multicolumn{3}{|c|}{Scenarios:} \\
 \hline
 Variable&No Bias&Bias\\
 \hline
 $\zeta$   & 0.3    &0.4\\
 $\beta$&   0&0\\
 \hline

\end{tabular}
 \caption{Values used in the simulation for Figure \ref{fig:typeone}. The presence of bias is varied by manipulating the underlying distribution of $Z$ in the external control group in comparison with the underlying distribution of $Z$ in the RCT group}
\end{table}

\subsubsection{Inflated type I error rates are corrected with the sensitivity analysis.}
\begin{figure}[H]
    \centering
    \includegraphics[width=1\textwidth]{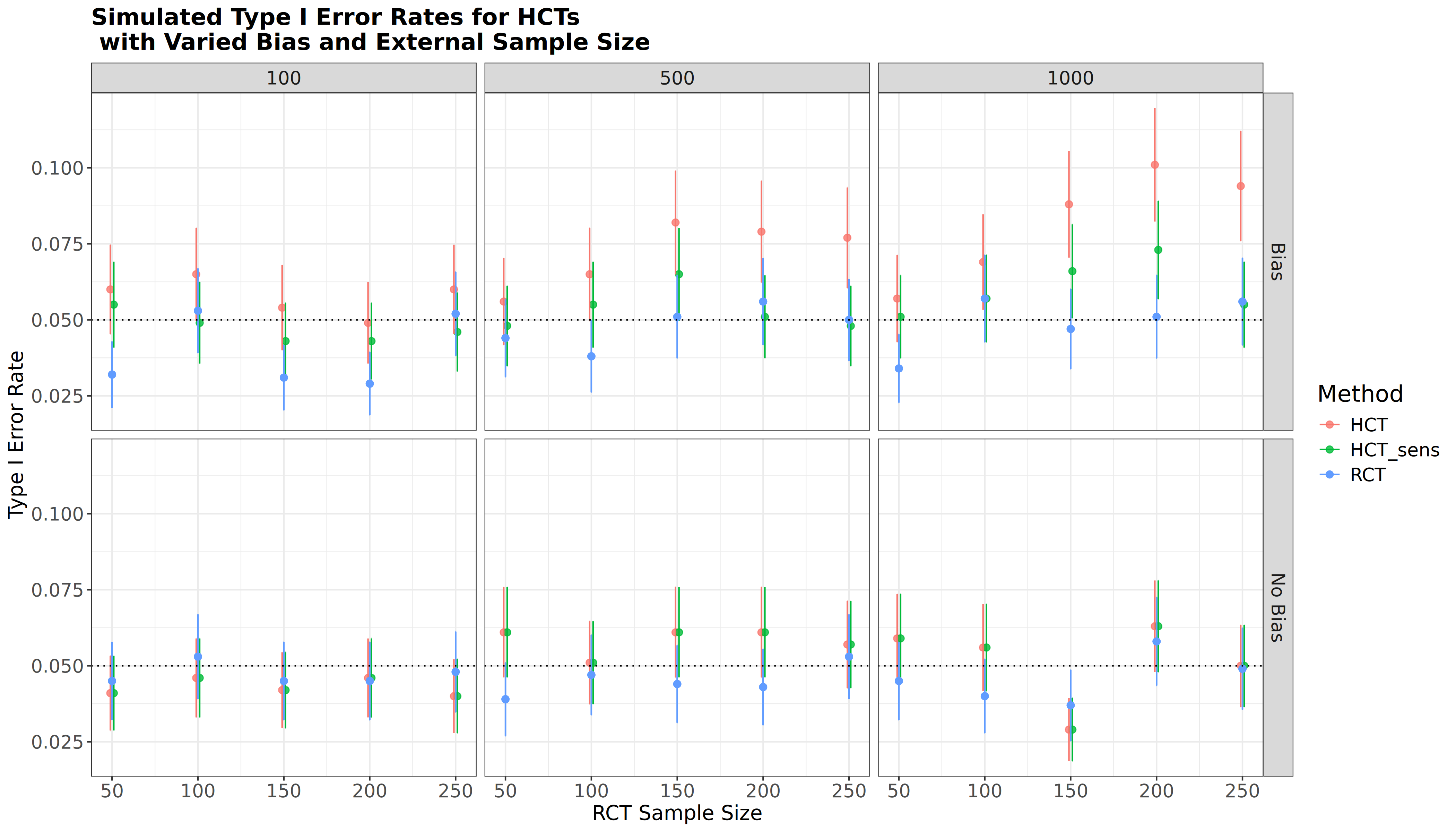}
    \caption{Type I error rates are compared across three trial setups: pure RCT, HCT, and HCT combined with the sensitivity analysis (shown in blue, red, and green respectively). Multiple scenarios are tested, varying the presence of bias and external sample sizes. From left to right: external sample size of 100, external sample size of 500, and external sample size of 1000. From top to bottom: bias present, no bias present. RCT sample sizes are varied with fixed external sample sizes.}
    \label{fig:typeone}
\end{figure}

Focusing first on the scenario where there is no bias, the sensitivity analysis returns the same results as the naive HCT. Since there is no bias, $B=0$ and calculations for confidence bounds return the same intervals as calculations for confidence intervals for naive HCTs. Here we can check that the coverage for the HCT trial-specific estimator is similar to that for the RCT ATE estimator without need to worry about the influence of bias. Despite no bias, the type I error rate of HCT analysis is slightly inflated when there are larger external sample sizes of 500 and 1000. We believe this reflects the nature of the second-order remainders of both since they are different between the trial-specific ATE and the RCT ATE. In Appendix \ref{2tsate} we provide a theoretical explanation for why type I errors of the trial-specific estimator may inflate when the ratio of external to trial data is large. However, as reflected in Figure \ref{fig:typeone}, these differences tend to disappear as RCT sample sizes increase.

When bias is present (and thus the sensitivity analysis is applied), we see that as the HCT makeup becomes more imbalanced, type I error rates tend to increase in a naive HCT analysis as seen in the red trend upwards across the panels. This makes sense considering that the influence of the biased external data is increasing with the imbalances. However, we see that the type I error rate is lowered within each panel as seen by the green values (HCT + sens) approaching the blue values (RCT only). With the sensitivity analysis, type I error control is maintained with error rates comparable to the HCT type I error rates in the unbiased scenario. In this way, the sensitivity analysis seems to account for the presence of bias, correcting the inflated type I error rates of naive HCT analysis and pulling them closer to the RCT type I error rates where $95\%$ coverage is approximated. Despite the improvements from the sensitivity analysis, we do see that the type I error rates are most similar to RCT type I error rates when external sample size is 100 or 500 -- when the external and RCT data is moderately balanced. In Appendix \ref{ticf2}, we modify the debiased machine learning cross-fitting to use two repetitions instead of one, resulting in significantly improved type I error rates. Due to this change, it is likely that the imbalances in the HCT data with large external data size result in variation from finite sample splitting, leading to the discrepancies seen in Figure \ref{fig:typeone} \cite{chernozhukov2024doubledebiasedmachinelearningtreatment}. In the case of heavily imbalanced HCT data, researchers may benefit from using two repetition cross-fitting to minimize finite sample splitting bias and maintain low type I error rates.

\subsection{Gains in Efficiency}

In the final figure, simulation is used to explore gains in trial efficiency through the use of power curves. Due to the interest in power, $\beta$ is set to 0.6 to simulate a true treatment effect. Similar to Figure \ref{fig:bias}, $q$ is varied as a function of the same desired external sample sizes and RCT sample sizes. The scenarios of `Favorable Bias', `Unfavorable Bias', and `No Bias' are explored with varying values of $\zeta$. In this case, bias has types of favorability based on whether it biases the estimate towards or away from an estimated null treatment effect. Again, the bias scenarios rely on the assumption that any deviations in the covariate distribution of $X$ are weak considering that the preliminary assessment of the external data should identify strong incompatibilities with the RCT population. Three analysis methods are compared: pure RCT analysis (with no external controls), naive HCT analysis, and HCT analysis combined with the sensitivity analysis. Type II errors were determined using the $95\%$ confidence regions for the HCT + sens. method and $95\%$ confidence intervals for the pure RCT and naive HCT analysis. Simulations with 1000 iterations were used to estimate the type II error rates. 
\newpage
\begin{table}[h]
\centering
\begin{tabular}{ |p{3cm}||p{3cm}|p{3cm}|p{3cm}|  }
 \hline
 \multicolumn{4}{|c|}{Scenarios:} \\
 \hline
 Variable&No Bias&Favorable Bias&Unfavorable Bias\\
 \hline
 $\zeta$   & 0.3    &0.4&   0.2\\
 $\beta$&   0.6  & 0.6      &0.6\\
 \hline
\end{tabular}
\caption{Values used in the simulation for Figure \ref{fig:enter-label}. The favorability of bias is varied by manipulating the underlying distribution of $Z$ in the external control group in comparison with the underlying distribution of $Z$ in the RCT group}
\end{table}

\subsubsection{Hybrid controls can be used to mitigate underpowering of small RCTs even in the presence of bias}
\begin{figure}[H]
    \centering
    \includegraphics[width=1\textwidth]{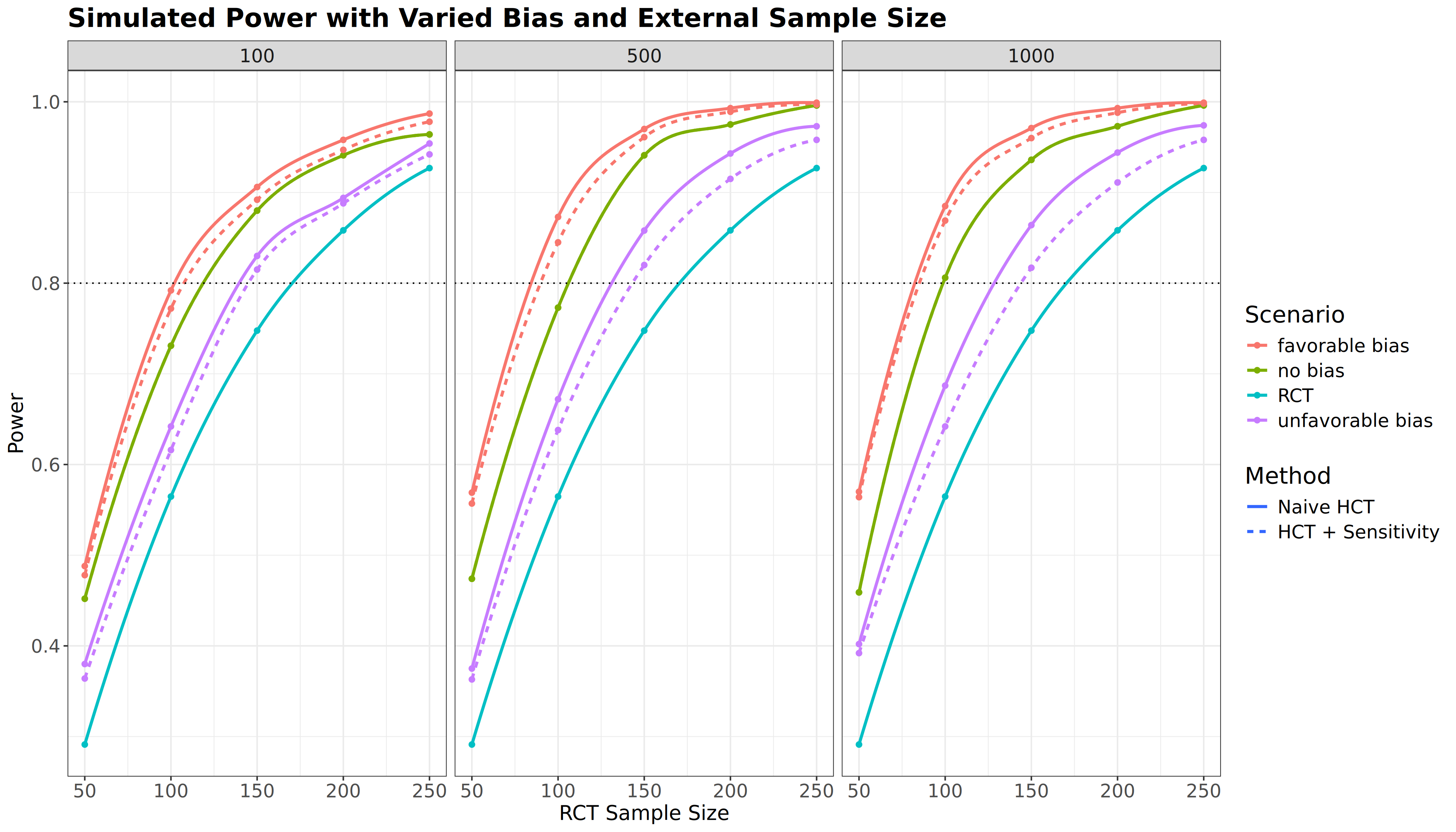}
    \caption{Power curves are compared across analyses based on pure RCT analysis (shown in solid blue lines), naive HCT analysis (shown in other colored solid lines), and HCT analysis combined with the sensitivity analysis (shown in dotted lines). External sample size is fixed in panels while RCT sample size is varied. From left to right: external sample size of 100, external sample size of 500, and external sample size of 1000. Bias scenarios are varied based on favorable bias, unfavorable bias, and no bias (shown in red, purple, and green).}
    \label{fig:enter-label}
\end{figure}

Simulations show that even in the presence of bias, HCTs can improve trial efficiency compared to underpowered RCTs. Across all simulated scenarios, power is always improved when an RCT is supplemented with external data. Note that this may not always be the case, especially if the external data is more different from the RCT control data than simulated, in which case it is possible that any gains in efficiency will be negated by large biases. However, in scenarios where the bias bound $B$ is estimated to be relatively large to the point that there is little benefit to using external controls, the contour plots and critical causal gaps described in Section \ref{sec3} would likely return results that advise researchers to proceed with a pure RCT analysis. As a result, it is unlikely that there would be many scenarios in which the sensitivity analysis results in worsened underpowering of trials (as researchers would simply use the RCT analysis, resulting in the same power). We also see that the amount of external data is also important as the dotted lines pull further away from their respective solid lines as the external sample size grows, aligning from findings from figure \ref{fig:bias} where larger external sample sizes result in an increase in $b$. However, in our simulated scenarios it seems as though this is usually offset by the gains in efficiency from the external data. 

Focusing on the sample size needed to reach $80\%$ power, it can be seen that roughly 170 participants are needed for a pure RCT analysis. However, using an HCT combined with sensitivity analysis, this number is roughly halved in the best simulated scenario (1000 external sample size with favorable bias). This reduction varies throughout scenarios depending on the direction of bias and ratio of data, but there was always at least a $10\%$ reduction in the required RCT sample size (smallest reduction when external sample size is 100 or 1000 with unfavorable bias). There seems to be a benefit to increasing the size of external controls from 100 to 500, but efficiency only slightly increases or slightly decreases when increasing the external controls from 500 to 1000 (a result of the increase in size of bias and standard error of the bias bound as seen in Section \ref{sec:biasbound}). Therefore, it is not crucial to have large amounts of external data. In fact, even when the external sample size is only 100, there are already visible gains in efficiency. This is especially important when considering rare disease or small population research where data is already scarce, potentially limiting the amount of available external data.

\subsection{Conclusion}
These results suggest that the use of external controls combined with the sensitivity analysis is a practical solution in relieving traditional RCT trial constraints, many of which stem from sample size requirements, even if the unrealistic mean exchangeability assumption is not met. These results also address and remedy the traditional complaints against naive HCTs where biases can greatly influence estimation and inference, notably inflating type I error rates.

\section{Demonstration}
\label{sec5}
In this section, we demonstrate how one can use external control data to supplement an RCT and subsequent sensitivity analysis to evaluate robustness of findings. Data for this case study was provided by Novo Nordisk A/S. We specifically focus on the analysis of the phase IIIb clinical trial NN9068-4229. The trial recruited participants who were inadequately controlled (defined as having a HbA1C of $7.0 - 11.0 $ \%) on treatment SGLT2i, a type of oral anti-diabetic treatment (OAD). The aim of the trial was to compare glycemic control of insulin IDegLira (treatment) against insulin IGlar (active control) as add-on therapy to SGLT2i in people diagnosed with Type 2 diabetes. The trial was a 26 week, 1:1 randomized, active-controlled, open label, treat-to-target trial with 420 enrolled participants and 419 trial participants. The outcome of the original analysis was the difference in change from baseline HbA1c to landmark visit week 26. In this case study, the primary objective is a binary endpoint Yes/No for being a responder i.e. having HbA1C level below $7.0 \%$ at the end of the trial (week 26). Since the original RCT study had significant findings of a positive treatment effect, we restrict the sample size to be smaller to more closely mirror a situation where an HCT may be desired. Specifically, we randomly sample 50 participants from each treatment arm, resulting in an RCT sample size of 100 with balanced treatment. External data originates from other previously conducted RCTs with study population consisting of people with Type 2 diabetes, who were inadequately controlled on their current OADs and receiving insulin IGlar. The external data originally consisted of $3,311$ participants, prior to selection of suitable external controls. See \cite{liao_prognostic_2025} for detailed information on the covariate distributions in the external data and RCT trial.

To assess the external data for suitability as external control data, we focused on two criteria: sufficient overlap of the covariates and mean exchangeability. Although the bias-adjusted inference from our method is robust against such violations, simulations and theory show that sufficient bias can invalidate significant results. Therefore, careful selection of external control data to avoid inducing unnecessary biases aids in the goal of increasing efficiency and maintaining correct significant findings. 

First, the baseline covariate distributions were compared between the external data participants and RCT participants. Baseline covariates included demographic information, laboratory measures such as baseline HbA1C, concomitant medication and vital signs. Although the continuous covariates were similar in distribution, some categorical covariates differed between the study populations. Due to concern of covariate overlap, select observations were excluded from the external data to ensure that all values of categorical covariates were observed in both the RCT and external data; for example, the titration target was restricted to be the same between the RCT and external arms since some external trials had different targets. Other covariates with complete lack of overlap included certain medication usage such as Alpha Glucosidase inhibitors. Selection of external controls in this manner is valid in that decisions are made based on a-priori knowledge of baseline covariates; further, this helps minimize population differences between the two arms. This process ultimately resulted in 878 external controls, resulting in a sample size of 978 for the HCT. 

Next, a test of mean exchangeability was conducted on the HCT data using a logistic model such as in Li et. al \cite{li_improving_2023}. Although the sensitivity analysis is built to measure violations of the mean exchangeability assumption, we believe that it remains valuable to examine potential violations. Biased external data can ultimately overwhelm the trial data, so testing for mean exchangeability can help us detect substantial biases in advance. To test the mean exchangeability assumption, the outcome $Y$ was regressed on $D$, the set of covariates $X$, and all interaction effects between $D$ and $X$ using only the control data. A chi-squared likelihood test was employed to test the null hypothesis that the coefficient of $D$ and the coefficients of all the interaction effects are equal to 0. No obvious violation of the mean exchangeability assumption was detected -- although there may be minor violations not detected by this parametric test -- so we proceed with the HCT analysis. Note that there are other tests of mean exchangeability that do not require parametric assumptions, but we use this one to mirror Li et al. \cite{li_improving_2023} and for simplicity of demonstration.

The average treatment effect was first calculated using just the RCT data, and then calculated using the full HCT data, both with debiased machine learning. 2 repetition 5 fold cross-fitting was used to mitigate finite sample splitting, which may result from the imbalance of the RCT and external control data. Point estimates of the trial-specific average treatment effect are reported in Table~\ref{tab:RCT_HCT}. Note that these do not account for any bias and do not rely on any of the sensitivity analysis methods; in essence, they are the estimates under the assumption that the mean exchangeability assumption is truly met. Using just the RCT data, the average treatment effect is positive, but not significant under a critical value of $0.05$. In the HCT analysis, the average treatment effect is likewise positive, but the results are significant under a critical value of $0.05$. The robustness value of the HCT analysis was $0.016$, indicating that if both $C^2_Y$ and $C^2_D$ are $0.016$ or greater, significant findings from the HCT will be negated, setting $\rho=1$ for a conservative estimate.

\begin{table}[h]
\centering
\begin{tabular}{ |p{3cm}||p{3cm}|p{3cm}|p{3cm}|  }
 \hline
 Method&Estimate&Standard Error&P-value\\
 \hline
 RCT   &  0.081    & 0.047 &   0.083\\
 HCT   &   0.103  & 0.049      &0.035\\
 \hline
\end{tabular}
\caption{Naive estimates of the trial-specific average treatment effect for the HCT and RCT samples, unadjusted for any potential biases.}\label{tab:RCT_HCT}
\end{table}

Benchmarking was then done using the covariates in a leave-one-out approach. For each covariate $X_j$, the HCT was re-analyzed excluding $X_j$ to get estimates of their corresponding sensitivity parameters, denoted by $\hat C^2_{Yj}$, $\hat C^2_{Dj}$, and $\hat \rho_j$. Functional components were fit for the `long' regressions, where $\tilde X$ includes $X_j$, and `short' regressions, where $X_{-j}$ denotes the set of covariates that excludes $X_j$. Fits of $\tilde \alpha(A, \tilde X, D)$, $\alpha(A,X_{-j},D)$, $\tilde \mu(A, \tilde X)$, and $\mu(A, X_{-j})$ were then plugged into the formulas from Section \ref{sec3.2.3} to obtain sample estimates of their corresponding sensitivity parameters. Results for the covariates with the three highest gains in explanatory power in each the outcome regression and Riesz Representer calculations are included in the tables below. These values will be used to assess the robustness of the HCT against hypothetical unmeasured confounders assuming that unmeasured confounders have less explanatory power than, or behave similarly to, the most explanatory measured covariates. This relies on the assumption that the measured covariates are thoughtfully chosen and measure important prognostic relationships.

\begin{table}[h]
\centering
\begin{tabular}{ |p{5cm}||p{2cm}|p{2cm}|p{2cm}|  }
 \hline
 Covariate&$\hat C^2_Y$&$\hat C^2_D$&$\hat \rho$\\
 \hline
 Baseline HbA1C   &  2.731    & 0.000&   0.000\\
 No Combination of Blood Glucose Lowering Drug  &   0.015  & 0.006      &-0.235\\
 Region: North America & 0.013 & 0.005 & -0.067 \\
 \hline
\end{tabular}
\caption{Top 3 most explanatory covariates in terms of $C^2_Y$ and corresponding estimates for $C^2_D$ and $\rho$.}\label{tab:cy}
\end{table}

\begin{table}[h]
\centering
\begin{tabular}{ |p{5cm}||p{2cm}|p{2cm}|p{2cm}|  }
 \hline
 Covariate&$\hat C^2_Y$&$\hat C^2_D$&$\hat \rho$\\
 \hline
 SGLT2i Continuation   &  0.000    & 0.172&   0.000\\
 Race: Nonwhite   &   0.002  & 0.082      & -0.343\\
 Baseline Sodium (mmol/L) & 0.010 & 0.075 & -0.515 \\
 \hline
\end{tabular}
\caption{Top 3 most explanatory covariates in terms of $C^2_D$ and corresponding estimates for $C^2_Y$ and $\rho$. }\label{tab:cd}
\end{table}

The covariate with the strongest explanatory power for the outcome regression is baseline HbA1C measure with a remarkably high value of 2.731 well above the robustness value. However, since the outcome of interest depends on the final HbA1C measure, this value is reasonable as people with higher values of HbA1C will be less likely to meet the cutoff HbA1C value of 7 \% and vice versa. Due to the strong mechanistic nature of the relationship here, it is doubtful that any unobserved covariate will have a relationship with the outcome as strong as baseline HbA1C. Further, the gain in explanatory power is restricted to the outcome regression; with 0 gain in explanatory power of the Riesz Representer, the resulting bias bound will be 0. Similarly, the covariate with the highest value of $\hat C^2_D$, SGLT2i continuation, is restricted to the Riesz Representer with 0 explanatory power in the outcome regression. It is also not surprising that this covariate has high explanatory power since it was part of the inclusion criteria of the RCT; the external data consists of patients on any OAD, which may be but is not restricted to SGLT2i. For these reasons, the sensitivity analysis will not use these covariates to obtain potential values of $C^2_Y$, $C^2_D$, and $\rho$; it is unlikely an unmeasured confounder will have a similar relationship with the outcome regression or the Riesz Representer, and even if so, the resulting bias is zero. However, it is a nice sanity check for assessing the performance of the benchmark process, given we a-priori know that these covariates will have strong explanatory power.

Based on the robustness value of 0.016 and the sensitivity parameters from the benchmark covariates, it does not seem likely that an unobserved confounder will have gains in explanatory power of both the outcome regression and Riesz Representer of magnitude 0.016 or higher. In fact, the highest observed gain in explanatory power of the outcome -- barring Baseline HbA1C measurement -- is 0.015. However, Table~\ref{tab:cd} shows that the robustness value is surpassed in $\hat C^2_D$ with `Race: Nonwhite' having a value of 0.082 well above the robustness value. Therefore, it is plausible that an unobserved confounder may increase the explanatory power of the Riesz Representer by more than 0.016. Further consideration is needed under different hypothetical combinations of $C^2_D$ and $C^2_Y$ as a high value of $C^2_D$ can invalidate significant findings even under low values of $C^2_Y$. 

For the sensitivity analysis we chose to use the estimated sensitivity parameters of the covariates `No Combination of Blood Glucose Lowering Drugs', `Race: Nonwhite', and `Baseline Sodium (mmol/L)'. We excluded `Region: North America'for brevity since we can see in Table \ref{tab:cy} that `No Combination of Blood Glucose Lowering Drugs' returned higher estimates of all three sensitivity parameters. Thus, focusing on the worst case scenario, we know that the sensitivity parameter estimates corresponding to `No Combination of Blood Glucose Lowering Drugs' will result in higher bias than `Region: North America'. In contrast, we chose to use both `Race: Nonwhite' and `Baseline Sodium (mmol/L)' from Table \ref{tab:cd} since neither dominated in all estimates of the sensitivity parameters. 

Confidence bounds are constructed as detailed in Section \ref{sec3} to obtain 95 \% confidence regions, using the estimated values of $C^2_D$ and $C^2_Y$ from the three benchmarking covariates. We constructed confidence bounds for each covariate under three different scenarios of $\rho$: most theoretically conservative value of $\rho=1$, most plausibly conservative, and benchmarked estimate of $\rho$ from the respective covariates. The most plausibly conservative value of $\rho$ was defined as $\max_j |\hat \rho _{j}|$ (the highest magnitude of $\hat \rho$ from the observed covariates), which incidently coincided with `Baseline Sodium (mmol/L)'. We choose to assess the bounds under multiple values of $\rho$ as it is the least intuitive sensitivity parameter. Recall that it is defined as the degree to which confounding in the outcome regression is related to confounding in the Riesz Representer while $C^2_Y$ and $C^2_D$ are the gains in explanatory power of the outcome regression and Riesz Representer, respectively, from a confounder. Therefore, if a variable has high values of all three, we can interpret it as having a large joint impact on the outcome and trial participation. On the other hand, if a variable has $\rho = 0$, a high value of $C^2_Y$ and $C^2_D=0$, it only impacts the outcome and adding external controls does not introduce omitted variable bias. Similarly, if a variable has $\rho = 0$, a high value of $C^2_D$ and $C^2_Y=0$, it only impacts trial participation but does not influence the outcome, again not introducing omitted variable bias.

\begin{table}[h]\label{tab:bench}
\centering
\begin{tabular}{ |p{6cm}||p{3cm}|p{3cm}|p{3cm}|  }
 \hline
 Benchmark Covariate&$\rho = 1$& $\rho =\max_j |\hat  \rho_{j}|$ & $\rho=\hat \rho_{j}$\\
 \hline
 No Combination of Blood Glucose Lowering Drugs (\%)   &  (0.009,0.197)  & (0.013,0.194) &(0.016,0.190)\\
 Race: Nonwhite & (0.004,0.202) & (0.010, 0.197) & (0.012,0.194)\\
 Baseline Sodium (mmol/L) & (-0.017,0.223) &(-0.006,0.212) &(-0.006,0.212)\\
 \hline
\end{tabular}
\caption{95 \% confidence regions constructed with one-sided 90\% confidence bounds, using `No Combination of Blood Glucose Lowering Drugs (\%)', `Race: Nonwhite', and `Baseline Sodium (mmol/L)'. Calculations used the benchmarked estimates of $C^2_D$ and $C^2_Y$ and varied values of $\rho$.}\label{tab:ci}
\end{table}

Table~\ref{tab:ci} reveals valuable information about the robustness of the HCT that the robustness value by itself did not. With the confidence bounds, we can see that there is a risk of bias that can invalidate the significant findings. Although the confidence bounds for `No Combination of Blood Glucose Lowering Drugs (\%)' and `Race: Nonwhite' remain statistically significant, we see that the bounds for `Baseline Sodium (mmol/L)' include zero. If we believe potential confounding variables to behave similarly to the first two covariates, we can be reasonably confident in the HCT findings. However, if we have reason to believe that a confounding variable may behave similarly as `Baseline Sodium (mmol/L)', we can no longer be confident in the HCT findings. This invites careful consideration of potential confounders as well as their potential joint impact on the outcome and Riesz Representer, discussed later in this section. This also highlights the importance of considering multiple covariates when benchmarking; if we had only used the covariates with the highest respective estimates of $C^2_D$ and $C^2_Y$, we would have missed this key insight. Again, the crucial factor with the sensitivity analysis is the \textbf{joint} effect of confounding on both outcome and trial participation (and potentially treatment assignment), not the isolated effect on one or the other.       

A contour plot was also created to visualize these findings, shown in Figure 5.  Since the point estimate is positive, only the contour plot for the lower bound is reported. Including the plot for the upper bound does not add to the discussion of statistical significance since we know the bound will be positive. The scenario for reference uses the benchmarked values of `Baseline Sodium (mmol/L)' ($\hat C^2_Y = 0.010$, $\hat C^2_D = 0.075$, and $\hat \rho = -0.515$); note that the values of $C^2_Y$ and $C^2_D$ given $\rho$ does not change the overall structure of the contour plot, just the reference point. Another contour plot using $\rho = 1$, the most conservative scenario, is also reported in Appendix \ref{app-rho1}. The reference point lies outside the statistically significant region (above the dotted red line), mirroring the finding from Table 9 that these values will result in statistically insignificant findings. Although the estimate for the lower bound is negative, it is near the line of significance. Therefore, if there is a confounder that behaves similarly but with slightly less explanatory power on the Riesz Representer, the results would be significant. Of course, if a confounder behaves similarly, but with more extreme confounding in either the outcome or the Riesz Representer, results would be insignificant. Note that any of the other covariates we used for the confidence bounds would have a reference point under the red line. Given these results, it is likely that one could still argue that the treatment insulin IdegLira has a positive treatment effect on being a responder (sufficient glycemic control) using the reduced sample RCT (with only 100 participants), which was not powered to this analysis. However, one must carefully consider the plausibility of such confounders since the sensitivity analysis shows some susceptibility to confounding bias.

\begin{figure}
    \centering
    \includegraphics[width=1\textwidth]{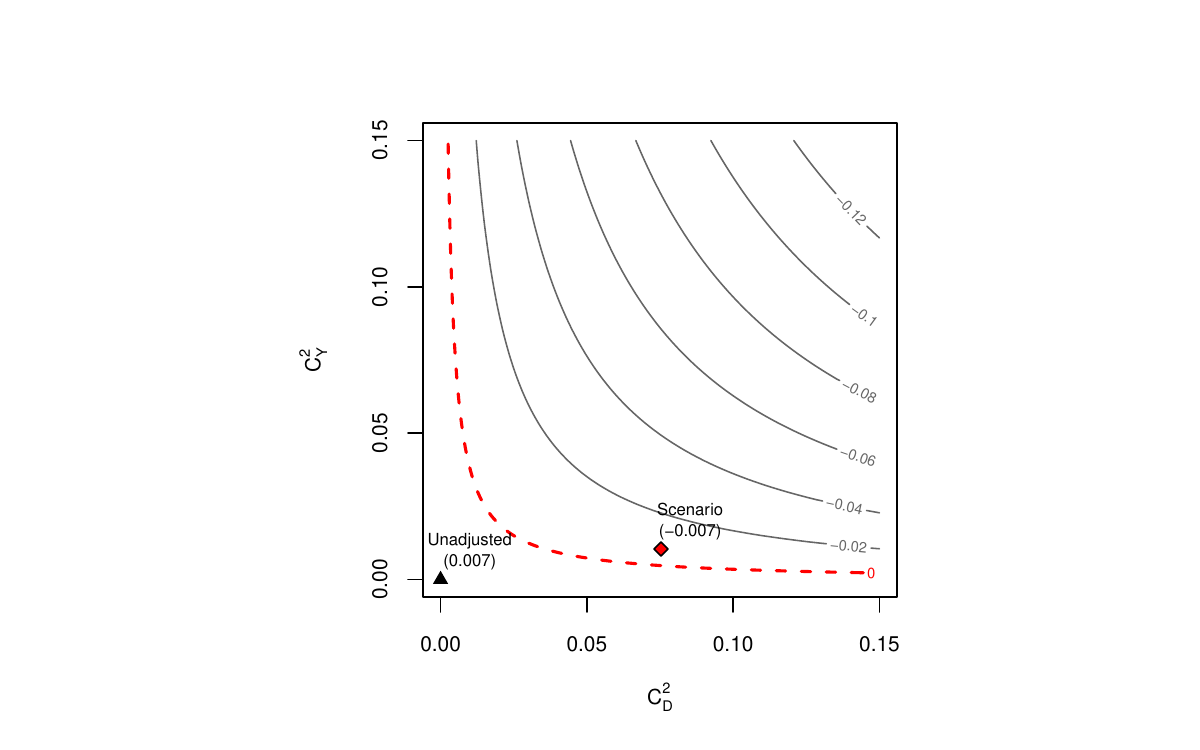}
    \caption{Sensitivity contour plot for the HCT. The unadjusted estimate of the lower bound as well as the adjusted estimate using the benchmark sensitivity values from `Baseline Sodium (mmol/L)' are reported. The red dotted line represents the point at which results are no longer statistically significant under $|\rho| = 0.515$, obtained from `Baseline Sodium (mmol/L)'.}
    \label{fig:placeholder}
\end{figure}

Finally, we must consider what potential confounders may be and whether we believe that they will behave similarly to the observed covariates. These may include factors that are not easily captured by blood tests and questionnaires that could reasonably be expected to affect the outcome (or differences in the RCT and external control arm populations) such as socio-economic status, environmental exposure (e.g. exposure to particulate pollution), detailed diet and exercise habits, mood/motivation, and adherence to treatment. In an RCT, we typically rely on randomization to balance these for unbiased estimation for a causal effect. However, in the HCT setting, we can no longer rely on randomization and our study more closely resembles an observational study. 

Comparing the effects of these variables with the observed covariates relies heavily on professional knowledge of the situation. For example, one can argue that no potential confounders would have a stronger effect on HbA1C levels than the measured covariates, given thoughtful selection of prognostic covariates. Further, one can argue if the effects of the unmeasured variables may be captured partially in the observed variables. It may be plausible that the effects of diet and exercise habits are partially captured within some of the baseline measurements from blood tests. On the other hand, it may not be plausible that the effects of mood/motivation are captured in the observed variables. 

Overall, the sensitivity analysis suggests that the results are most susceptible to confounding through the Riesz Representer (i.e. trial participation and treatment allocation). When discussing the effect of these potential confounders on the Riesz Representer, one must consider how the external data was collected and how the population may differ from the RCT population. Careful selection of external controls can increase the robustness of the HCT against confounding bias if they are selected such that it is unlikely that any unobserved covariates can have a strong influence on trial participation. In this case, we should ask if there are any confounders that may be significantly different between the external and RCT population since the benchmark results revealed that the greatest susceptibility to bias lies in $C^2_D$. Ultimately, the sensitivity analysis is a tool for measuring overall robustness against confounding in HCTs, but professional knowledge is required to assess the assumptions that are used for benchmarking.

\section{Discussion}
\label{sec6}
The findings in Section \ref{sec4} imply that there needs to be careful deliberation in selecting the ideal ratio of RCT data against the external control data. From Figure \ref{fig:typeone}, we see finite sample size issues with type I error inflation when data are heavily imbalanced, even when there is no bias present. This finding stems from the form of the second-order remainder of the trial-specific ATE as described in Appendix \ref{2tsate} where it takes longer for the asymptotics to `kick in' compared to the RCT ATE when $\hat q$ is small.  We believe this to be an important finding, as the form of the remainder has not been explored in prior trial-specific ATE literature, despite its relevance to understanding the asymptotic behavior of estimators under certain conditions, such as varying values of $q$. However, regardless of the external sample sizes in the simulations, these issues seem to disappear once the RCT sample size reaches 100. As a result, we recommend being wary of using unbalanced ratios when RCT sample size is especially small. However, as seen in Figure \ref{fig:enter-label}, efficiency can be safely improved through the combined use of external data and sensitivity analysis even when using smaller external sample sizes. 

When RCT sample size is sufficiently large, the possibility of using larger external sample sizes arises. The question of how large depends on several factors. First, we must consider the expected effect size of treatment, and then we must consider how different the external control arm may be from the RCT control arm. However, this second consideration should not be a major concern given proper selection of the external control arm. In fact, it would be improper to proceed with HCT analysis if the external control arm were expected to be significantly different than the RCT control arm. Despite this, it is worth considering the size of the inevitable bias from confounding since it can be seen in Figures \ref{fig:bias} and \ref{fig:enter-label} that any difference is amplified when external sample size is increased. The expected effect size is then important here; if the expected effect size is relatively small, amplified bias threatens correctly identifying a treatment effect. In this case, we recommend using a smaller external control arm. However, in the case that the expected effect size is relatively large (such as in the simulations), we see in Figure \ref{fig:enter-label} that we can safely increase power by increasing the external sample size. The greatest power gains were achieved in the simulations when external sample size was 500 or 1000. While the gains in power are large from 100 to 500, the gains in power are relatively smaller from 500 to 1000. Consequently, it may be best to still limit the ratio of data away from extreme imbalances since the potential consequences outweigh the marginal benefits of relying heavily on external data.   

These considerations, however, all rely on the assumption made in the simulations that the true values of $C^2_Y$ and $C^2_D$ are known. In reality, the size of confounding due to omitted variable bias is never truly known. As mentioned above, omitted variable bias and sensitivity literature has explored different benchmarking algorithms to determine the relative plausible sizes of unobserved confounding \cite{altonji_selection_2005, chernozhukov_long_2024, cinelli_making_2020, imbens_sensitivity_2003, bonvini_sensitivity_2022}. In particular, benchmarking methods are popular where the observed covariates are used to approximate the influence of unobserved confounders. It has been argued that given strategic selection of measured covariate variables, the unobserved covariates likely are equally or less explanatory of outcome and treatment assignment \cite{altonji_selection_2005}. With this assumption, one can use observed covariates to estimate bounds on the relationships of any unobserved variables with outcome and treatment. \citet{mcclean_calibrated_2024} summarizes these methods where notable strategies include the ``leave-one-out" approach, addressed earlier in the methods section, as well as more complicated methods such as ``leave-some-out" and ``leave-every-out" to also rely on subsets of observed covariates. In all, these algorithmic approaches hold promise as benchmarking methods and can be used in combination with this sensitivity analysis to estimate plausible ranges of values for $C^2_Y$ and $C^2_D$ in order to judge robustness against omitted variable bias. While benchmarking itself provides measures of confounding due to unobserved confounders, critics have pointed out that relying solely on benchmarking may lead to erroneous conclusions \cite{cinelli_making_2020, mcclean_calibrated_2024}. The benefit of using these benchmarks in conjunction with the sensitivity analysis outlined here is that the bias bound is properly identified, leading to safer conclusions on the causal effect of treatment \cite{cinelli_making_2020}. 

Finally, the scope of this paper focuses on binary outcomes for simplicity. However, we expect these findings to generalize easily to the continuous outcome scenario. The main motivation behind focusing on binary outcomes was the $r(X)$ term, the ratio of variance in the outcome between the RCT control data and external control data, in the Li estimator for the trial-specific average treatment effect. To reduce the number of values to be estimated, we binarized the outcome such that $r(X)$ would equal 1 under mean exchangeability. For continuous outcomes where $r(X)$ would need to be estimated, the original paper provides insight on empirical methods \cite{li_improving_2023}. Further, we believe that the results would translate to continuous outcomes since the variance ratio is expected to only affect efficiency, not bias. Therefore, the idea of increasing trial efficiency while protecting against bias with the sensitivity analysis should be able to be extended to continuous outcomes after some verification with simulations. 

For estimation, we relied on an efficient estimating equations approach with debiased machine learning. However, so long as the framework in the paper is followed, other estimation techniques can also be used. For example, methods such as targeted maximum likelihood estimation (TMLE), which also uses the efficient influence function, may be implemented \cite{chernozhukov_long_2024, laan_targeted_2011}. However, we relied on an EEE approach for an intuitively understandable approach by prospective researchers who may implement the sensitivity analysis. Further, we relied on the Li estimator for the trial-specific ATE due to its efficiency, but the sensitivity analysis can be applied for other estimators of the trial-specific ATE so long as the Riesz representer and outcome regression can be properly identified.  

For selection of the external control population, it is important that the external control subjects are as similar as possible to the RCT control subjects to maintain the integrity of results. Several guidelines exist for selection of external controls, such as Pocock's selection criteria \cite{pocock_combination_1976}. In particular, the external control subjects must have: 
\begin{enumerate}
    \item Received the same treatment as the randomized controls
    \item Been selected from a recent clinical study with comparable eligibility criteria
    \item Been evaluated in the same method as the randomized controls
    \item Had comparable distributions of important covariates as the randomized controls
    \item Been selected from a study conducted by the same organization with overlap in clinical investigators
    \item No indications of significant differences between the external and randomized controls.
\end{enumerate}

These strict criteria attempt to ensure that the external controls are as similar as possible to the randomized control subjects. However, it is argued in some literature that these can be relaxed in extreme cases where data is limited or there is a pressing need to conduct research \cite{lim_minimizing_2018}. For example, there has been movement towards also including electronic health record (EHR) data as a source of external control data \cite{carrigan}. The implication of adding EHR data as an external data source is large, vastly enriching the availability of recent, relevant health data. Regardless of a relaxation of these criteria, data selection should still mirror these criteria as closely as practically possible to balance minimizing biased decision making and minimizing the patient burden. Further, as discussed in Section \ref{sec5}, careful selection of the external control sample can help loosely control the magnitude of $C^2_D$. When population differences are smaller, the potential impact a confounder has on trial participation is likewise smaller. Although this method works even if the populations are dissimilar, it is important that researchers carefully select external control samples to avoid inducing unnecessary bias in estimation.

Other attempts to minimize bias from external control selection include a priori approaches such as a test-then-pool approach \cite{viele_use_2014}. In this approach, a simple hypothesis test is conducted on the outcome prior to the decision to pool the external and randomized controls. Given an insignificant finding, researchers pool the data together and analyze the data as an HCT. Given a significant finding, the hypothesis test raises a red flag, resulting in researchers analyzing the RCT separately from the external data. \citet{li_improving_2023} also suggest a test for the mean exchangeability assumption in which a parametric outcome model is fit with interactions between $D$ and $X$. A variety of other tests exist in which this mean exchangeability assumption can be tested in advance of pooling such as the non-parametric omnibus test of equality \cite{luedtke_omnibus_2019}  However, these approaches are all limited in that they are only asymptotically guaranteed to detect discrepancies in the data. Since there will always be a level of bias between the control groups, these tests will always choose to not pool the data asymptotically. However, these are applied to small samples in the HCT setting where the availability of data (from the RCT and/or external data source) is constrained. As a result, these will only identify any significantly large discrepancies in the data. The sensitivity analysis is thus a logical complement to such tests (and pooling) to detect the smaller inevitable differences in the data that lead to bias in naive HCT analysis.

The ability to safely use external control data in hybrid control trials has multiple implications. HCTs have been proposed as early as 1976, but this trial design has typically been discouraged by major regulatory bodies \cite{noauthor_international_2000}. Despite openness to the use of external data, the reliance on the mean exchangeability assumption, which practically is always violated to some degree, has hindered the wide acceptance of HCTs. HCT literature has predominantly focused on either creating new, more efficient estimators or relaxing the mean exchangeability assumption. For example, Bayesian methodologies have been proposed, in which one can dynamically adjust how much information to borrow from the external data based on compatibility of the controls, but gains in efficiency are minimal when enforcing strict type I error control \cite{koppschneider_power_2020}. Previous HCT literature focuses on mitigating biases through dynamic methods whereas we focus on quantifying bias under the worst case scenario. This paper fills in this gap in the research, quantifying this inevitable bias through a sensitivity analysis. We hope that filling in this gap in the research and making findings more credible will lead to the wider acceptance of HCTs. 

The benefits of the HCT trial design are numerous. Of course, there is reduced recruitment, leading to shorter timelines and budgets, but the benefits go beyond these factors. Ethically, the HCT trial design is attractive as it minimizes patient burden \cite{lim_minimizing_2018}. A study relying on solely an RCT requires more patients to be randomized, resulting in a larger number of patients missing out on receiving a treatment. In addition, with increased control information, it is possible to change the randomization ratio such that more RCT participants receive the treatment than the control. This not only is more ethical but can also be more attractive to prospective participants, encouraging trial participation. Another benefit of adding credibility to HCTs is that it can decrease reliance on single-arm study designs in which all control data come from external sources, where possible. Single-arm trials are also used in small sample settings, but rely on a stronger version of the mean exchangeability assumption, resulting in an even higher risk of bias \cite{tan_augmenting_2022}. While there are cases in which even an HCT may be infeasible, the cases in which an HCT is feasible should provide a safer alternative trial design than the single-arm trial \cite{lim_minimizing_2018}.

Considering its extreme nature, HCTs are only recommended in situations where there is a pressing need to borrow information from outside of an RCT. In particular, this includes rare disease research, oncology trials, and traditionally excluded populations from research. Due to the nature of these research areas, conducting a full-scale RCT is usually impractical, hindering the development of medicines and treatments for these populations. The findings of this paper align with the aims of the 21st Century Cures Act ``to accelerate the discovery, development, and delivery of 21st century cures" \cite{noauthor_21st_2017}. With added credibility to the findings from HCTs, regulatory bodies will hopefully allow for more HCT trials, greatly increasing the volume of research that can be dedicated to these traditionally difficult areas in medical research.

\bibliographystyle{plainnat}
\bibliography{bib.bib}

@misc{gao_improving_2024,
	title = {Improving randomized controlled trial analysis via data-adaptive borrowing},
	url = {http://arxiv.org/abs/2306.16642},
	doi = {10.48550/arXiv.2306.16642},
	urldate = {2025-07-10},
	publisher = {arXiv},
	author = {Gao, Chenyin and Yang, Shu and Shan, Mingyang and Ye, Wenyu and Lipkovich, Ilya and Faries, Douglas},
	month = nov,
	year = {2024},
	note = {arXiv:2306.16642 [stat]},
}

@article{sox_methods_2012,
	title = {The methods of comparative effectiveness research},
	volume = {33},
	issn = {1545-2093},
	doi = {10.1146/annurev-publhealth-031811-124610},
	language = {eng},
	journal = {Annual Review of Public Health},
	author = {Sox, Harold C. and Goodman, Steven N.},
	month = apr,
	year = {2012},
	pmid = {22224891},
	pages = {425--445},
}

@article{bentley_conducting_2019,
	title = {Conducting clinical trials-costs, impacts, and the value of clinical trials networks: {A} scoping review},
	volume = {16},
	issn = {1740-7753},
	shorttitle = {Conducting clinical trials-costs, impacts, and the value of clinical trials networks},
	doi = {10.1177/1740774518820060},
	language = {eng},
	number = {2},
	journal = {Clinical Trials (London, England)},
	author = {Bentley, Colene and Cressman, Sonya and van der Hoek, Kim and Arts, Karen and Dancey, Janet and Peacock, Stuart},
	month = apr,
	year = {2019},
	pmid = {30628466},
	pages = {183--193},
}

@incollection{glennerster_chapter_2017,
	series = {Handbook of {Field} {Experiments}},
	title = {Chapter 5 - {The} {Practicalities} of {Running} {Randomized} {Evaluations}: {Partnerships}, {Measurement}, {Ethics}, and {Transparency}},
	volume = {1},
	shorttitle = {Chapter 5 - {The} {Practicalities} of {Running} {Randomized} {Evaluations}},
	url = {https://www.sciencedirect.com/science/article/pii/S2214658X16300150},
	urldate = {2025-01-29},
	booktitle = {Handbook of {Economic} {Field} {Experiments}},
	publisher = {North-Holland},
	author = {Glennerster, R.},
	editor = {Banerjee, Abhijit Vinayak and Duflo, Esther},
	month = jan,
	year = {2017},
	doi = {10.1016/bs.hefe.2016.10.002},
	pages = {175--243},
}

@article{temple_placebo-controlled_2000,
	title = {Placebo-controlled trials and active-control trials in the evaluation of new treatments. {Part} 1: ethical and scientific issues},
	volume = {133},
	issn = {0003-4819},
	shorttitle = {Placebo-controlled trials and active-control trials in the evaluation of new treatments. {Part} 1},
	doi = {10.7326/0003-4819-133-6-200009190-00014},
	language = {eng},
	number = {6},
	journal = {Annals of Internal Medicine},
	author = {Temple, R. and Ellenberg, S. S.},
	month = sep,
	year = {2000},
	pmid = {10975964},
	pages = {455--463},
}

@article{jahanshahi_use_2021,
	title = {The {Use} of {External} {Controls} in {FDA} {Regulatory} {Decision} {Making}},
	volume = {55},
	issn = {2168-4804},
	doi = {10.1007/s43441-021-00302-y},
	language = {eng},
	number = {5},
	journal = {Therapeutic Innovation \& Regulatory Science},
	author = {Jahanshahi, Mahta and Gregg, Keith and Davis, Gillian and Ndu, Adora and Miller, Veronica and Vockley, Jerry and Ollivier, Cecile and Franolic, Tanja and Sakai, Sharon},
	month = sep,
	year = {2021},
	pmid = {34014439},
	pmcid = {PMC8332598},
	pages = {1019--1035},
}

@article{pocock_combination_1976,
	title = {The combination of randomized and historical controls in clinical trials},
	volume = {29},
	issn = {0021-9681},
	url = {https://www.sciencedirect.com/science/article/pii/0021968176900448},
	doi = {10.1016/0021-9681(76)90044-8},
	number = {3},
	urldate = {2025-01-29},
	journal = {Journal of Chronic Diseases},
	author = {Pocock, Stuart J.},
	month = mar,
	year = {1976},
	pages = {175--188},
}

@article{tan_augmenting_2022,
	title = {Augmenting control arms with real-world data for cancer trials: {Hybrid} control arm methods and considerations},
	volume = {30},
	issn = {2451-8654},
	shorttitle = {Augmenting control arms with real-world data for cancer trials},
	doi = {10.1016/j.conctc.2022.101000},
	language = {eng},
	journal = {Contemporary Clinical Trials Communications},
	author = {Tan, W. Katherine and Segal, Brian D. and Curtis, Melissa D. and Baxi, Shrujal S. and Capra, William B. and Garrett-Mayer, Elizabeth and Hobbs, Brian P. and Hong, David S. and Hubbard, Rebecca A. and Zhu, Jiawen and Sarkar, Somnath and Samant, Meghna},
	month = dec,
	year = {2022},
	pmid = {36186544},
	pmcid = {PMC9519429},
	pages = {101000},
}

@article{baumfeld_andre_trial_2020,
	title = {Trial designs using real-world data: {The} changing landscape of the regulatory approval process},
	volume = {29},
	issn = {1099-1557},
	shorttitle = {Trial designs using real-world data},
	doi = {10.1002/pds.4932},
	language = {eng},
	number = {10},
	journal = {Pharmacoepidemiology and Drug Safety},
	author = {Baumfeld Andre, Elodie and Reynolds, Robert and Caubel, Patrick and Azoulay, Laurent and Dreyer, Nancy A.},
	month = oct,
	year = {2020},
	pmid = {31823482},
	pmcid = {PMC7687110},
	pages = {1201--1212},
}

@misc{noauthor_international_2000,
	title = {International {Conference} on {Harmonization} ({ICH}) {E10}: {Choice} of control group and related issues in clinical trials.},
	url = {https://database.ich.org/sites/default/files/E10\_Guideline.pdf},
	month = jul,
	year = {2000},
}

@misc{noauthor_fda_2019,
	title = {{FDA} {Guidance} {For} {Industry}: {Rare} diseases–natural history studies for drug development.},
	url = {https://www.fda.gov/media/122425/download},
	month = mar,
	year = {2019},
}

@misc{noauthor_fda_2019-2,
	title = {{FDA} {Guidance} {For} {Industry}: {Demonstrating} substantial evidence of effectiveness for human drug and biological products.},
	url = {https://www.fda.gov/media/133660/download},
	month = dec,
	year = {2019},
}

@misc{noauthor_framework_2018,
	title = {Framework for {FDA}'s real-world evidence program},
	url = {https://www.fda.gov/media/120060/download},
	month = dec,
	year = {2018},
}

@misc{noauthor_21st_2017,
	title = {21st {Century} {Cures} {Act}},
	url = {https://www.gpo.gov/fdsys/pkg/BILLS-114hr34enr/pdf/BILLS-114hr34enr.pdf},
	month = feb,
	year = {2017},
}

@misc{lee2025rieszboostgradientboostingriesz,
      title={RieszBoost: Gradient Boosting for Riesz Regression}, 
      author={Kaitlyn J. Lee and Alejandro Schuler},
      year={2025},
      eprint={2501.04871},
      archivePrefix={arXiv},
      primaryClass={stat.ML},
      url={https://arxiv.org/abs/2501.04871}, 
}

@misc{intro,
    title= {Introduction to Modern Causal Inference},
    author={Alejandro Schuler and Mark van der Laan}
}

@misc{chernozhukov2024doubledebiasedmachinelearningtreatment,
      title={Double/Debiased Machine Learning for Treatment and Causal Parameters}, 
      author={Victor Chernozhukov and Denis Chetverikov and Mert Demirer and Esther Duflo and Christian Hansen and Whitney Newey and James Robins},
      year={2024},
      eprint={1608.00060},
      archivePrefix={arXiv},
      primaryClass={stat.ML},
      url={https://arxiv.org/abs/1608.00060}, 
}

@article{viele_use_2014,
	title = {Use of historical control data for assessing treatment effects in clinical trials},
	volume = {13},
	copyright = {Copyright © 2013 John Wiley \& Sons, Ltd.},
	issn = {1539-1612},
	url = {https://onlinelibrary.wiley.com/doi/abs/10.1002/pst.1589},
	doi = {10.1002/pst.1589},
	language = {en},
	number = {1},
	urldate = {2025-04-28},
	journal = {Pharmaceutical Statistics},
	author = {Viele, Kert and Berry, Scott and Neuenschwander, Beat and Amzal, Billy and Chen, Fang and Enas, Nathan and Hobbs, Brian and Ibrahim, Joseph G. and Kinnersley, Nelson and Lindborg, Stacy and Micallef, Sandrine and Roychoudhury, Satrajit and Thompson, Laura},
	year = {2014},
	note = {\_eprint: https://onlinelibrary.wiley.com/doi/pdf/10.1002/pst.1589},
	pages = {41--54},
}

@article{koppschneider_power_2020,
	title = {Power gains by using external information in clinical trials are typically not possible when requiring strict type {I} error control},
	volume = {62},
	issn = {0323-3847, 1521-4036},
	url = {https://onlinelibrary.wiley.com/doi/10.1002/bimj.201800395},
	doi = {10.1002/bimj.201800395},
	language = {en},
	number = {2},
	urldate = {2025-04-29},
	journal = {Biometrical Journal},
	author = {Kopp‐Schneider, Annette and Calderazzo, Silvia and Wiesenfarth, Manuel},
	month = mar,
	year = {2020},
	pages = {361--374},
}

@article{luedtke_omnibus_2019,
	title = {An omnibus non-parametric test of equality in distribution for unknown functions},
	volume = {81},
	copyright = {© 2018 Royal Statistical Society},
	issn = {1467-9868},
	url = {https://onlinelibrary.wiley.com/doi/abs/10.1111/rssb.12299},
	doi = {10.1111/rssb.12299},
	language = {en},
	number = {1},
	urldate = {2025-04-28},
	journal = {Journal of the Royal Statistical Society: Series B (Statistical Methodology)},
	author = {Luedtke, Alex and Carone, Marco and van der Laan, Mark J.},
	year = {2019},
	note = {\_eprint: https://onlinelibrary.wiley.com/doi/pdf/10.1111/rssb.12299},
	pages = {75--99},
}

@article{carrigan,
	title = {Using {Electronic} {Health} {Records} to {Derive} {Control} {Arms} for {Early} {Phase} {Single}‐{Arm} {Lung} {Cancer} {Trials}: {Proof}‐of‐{Concept} in {Randomized} {Controlled} {Trials}},
	volume = {107},
	issn = {0009-9236},
	shorttitle = {Using {Electronic} {Health} {Records} to {Derive} {Control} {Arms} for {Early} {Phase} {Single}‐{Arm} {Lung} {Cancer} {Trials}},
	url = {https://www.ncbi.nlm.nih.gov/pmc/articles/PMC7006884/},
	doi = {10.1002/cpt.1586},
	number = {2},
	urldate = {2025-04-28},
	journal = {Clinical Pharmacology and Therapeutics},
	author = {Carrigan, Gillis and Whipple, Samuel and Capra, William B. and Taylor, Michael D. and Brown, Jeffrey S. and Lu, Michael and Arnieri, Brandon and Copping, Ryan and Rothman, Kenneth J.},
	month = feb,
	year = {2020},
	pmid = {31350853},
	pmcid = {PMC7006884},
	pages = {369--377},
}

@article{imbens,
author = {Imbens, Guido W. and Manski, Charles F.},
title = {Confidence Intervals for Partially Identified Parameters},
journal = {Econometrica},
volume = {72},
number = {6},
pages = {1845-1857},
doi = {https://doi.org/10.1111/j.1468-0262.2004.00555.x},
url = {https://onlinelibrary.wiley.com/doi/abs/10.1111/j.1468-0262.2004.00555.x},
eprint = {https://onlinelibrary.wiley.com/doi/pdf/10.1111/j.1468-0262.2004.00555.x},
year = {2004}
}

@article{valancius_causal_2024,
	title = {A causal inference framework for leveraging external controls in hybrid trials},
	volume = {80},
	issn = {0006-341X},
	url = {https://doi.org/10.1093/biomtc/ujae095},
	doi = {10.1093/biomtc/ujae095},
	number = {4},
	urldate = {2025-01-29},
	journal = {Biometrics},
	author = {Valancius, Michael and Pang, Herbert and Zhu, Jiawen and Cole, Stephen R and Jonsson Funk, Michele and Kosorok, Michael R},
	month = dec,
	year = {2024},
	pages = {ujae095},
}

@book{laan_targeted_2011,
	title = {Targeted {Learning}: {Causal} {Inference} for {Observational} and {Experimental} {Data}},
	isbn = {978-1-4419-9781-4},
	author = {Laan, Mark and Rose, Sherri},
	month = jan,
	year = {2011},
	doi = {10.1007/978-1-4419-9782-1},
}

@misc{mcclean_calibrated_2024,
	title = {Calibrated sensitivity models},
	url = {http://arxiv.org/abs/2405.08738},
	doi = {10.48550/arXiv.2405.08738},
	urldate = {2025-05-21},
	publisher = {arXiv},
	author = {McClean, Alec and Branson, Zach and Kennedy, Edward H.},
	month = nov,
	year = {2024},
	note = {arXiv:2405.08738 [stat]},
}

@article{altonji_selection_2005,
	title = {Selection on {Observed} and {Unobserved} {Variables}: {Assessing} the {Effectiveness} of {Catholic} {Schools}},
	volume = {113},
	issn = {0022-3808},
	shorttitle = {Selection on {Observed} and {Unobserved} {Variables}},
	url = {https://www.jstor.org/stable/10.1086/426036},
	doi = {10.1086/426036},
	number = {1},
	urldate = {2025-05-21},
	journal = {Journal of Political Economy},
	author = {Altonji, Joseph G. and Elder, Todd E. and Taber, Christopher R.},
	year = {2005},
	note = {Publisher: The University of Chicago Press},
	pages = {151--184},
}

@article{bonvini_sensitivity_2022,
	title = {Sensitivity analysis via the proportion of unmeasured confounding},
	volume = {117},
	issn = {0162-1459, 1537-274X},
	url = {http://arxiv.org/abs/1912.02793},
	doi = {10.1080/01621459.2020.1864382},
	number = {539},
	urldate = {2025-05-21},
	journal = {Journal of the American Statistical Association},
	author = {Bonvini, Matteo and Kennedy, Edward H.},
	month = jul,
	year = {2022},
	note = {arXiv:1912.02793 [stat]},
	pages = {1540--1550},
}

@article{cinelli_making_2020,
	title = {Making {Sense} of {Sensitivity}: {Extending} {Omitted} {Variable} {Bias}},
	volume = {82},
	copyright = {https://academic.oup.com/journals/pages/open\_access/funder\_policies/chorus/standard\_publication\_model},
	issn = {1369-7412, 1467-9868},
	shorttitle = {Making {Sense} of {Sensitivity}},
	url = {https://academic.oup.com/jrsssb/article/82/1/39/7056023},
	doi = {10.1111/rssb.12348},
	language = {en},
	number = {1},
	urldate = {2025-05-21},
	journal = {Journal of the Royal Statistical Society Series B: Statistical Methodology},
	author = {Cinelli, Carlos and Hazlett, Chad},
	month = feb,
	year = {2020},
	pages = {39--67},
}

@article{duan_evaluating_2006,
	title = {Evaluating water quality using power priors to incorporate historical information},
	volume = {17},
	copyright = {http://doi.wiley.com/10.1002/tdm\_license\_1.1},
	issn = {1180-4009, 1099-095X},
	url = {https://onlinelibrary.wiley.com/doi/10.1002/env.752},
	doi = {10.1002/env.752},
	language = {en},
	number = {1},
	urldate = {2025-06-02},
	journal = {Environmetrics},
	author = {Duan, Yuyan and Ye, Keying and Smith, Eric P.},
	month = feb,
	year = {2006},
	pages = {95--106},
}

@article{agrawal_use_2023,
	title = {Use of {Single}-{Arm} {Trials} for {US} {Food} and {Drug} {Administration} {Drug} {Approval} in {Oncology}, 2002-2021},
	volume = {9},
	issn = {2374-2437},
	url = {https://jamanetwork.com/journals/jamaoncology/fullarticle/2800126},
	doi = {10.1001/jamaoncol.2022.5985},
	language = {en},
	number = {2},
	urldate = {2025-06-02},
	journal = {JAMA Oncology},
	author = {Agrawal, Sundeep and Arora, Shaily and Amiri-Kordestani, Laleh and De Claro, R. Angelo and Fashoyin-Aje, Lola and Gormley, Nicole and Kim, Tamy and Lemery, Steven and Mehta, Gautam U. and Scott, Emma C. and Singh, Harpreet and Tang, Shenghui and Theoret, Marc R. and Pazdur, Richard and Kluetz, Paul G. and Beaver, Julia A.},
	month = feb,
	year = {2023},
	pages = {266},
}

@article{li_revisit_2020,
	title = {Revisit of test‐then‐pool methods and some practical considerations},
	volume = {19},
	issn = {1539-1604, 1539-1612},
	url = {https://onlinelibrary.wiley.com/doi/10.1002/pst.2009},
	doi = {10.1002/pst.2009},
	language = {en},
	number = {5},
	urldate = {2025-06-02},
	journal = {Pharmaceutical Statistics},
	author = {Li, Wen and Liu, Frank and Snavely, Duane},
	month = sep,
	year = {2020},
	pages = {498--517},
}

@article{ghadessi_roadmap_2020,
	title = {A roadmap to using historical controls in clinical trials – by {Drug} {Information} {Association} {Adaptive} {Design} {Scientific} {Working} {Group} ({DIA}-{ADSWG})},
	volume = {15},
	issn = {1750-1172},
	url = {https://ojrd.biomedcentral.com/articles/10.1186/s13023-020-1332-x},
	doi = {10.1186/s13023-020-1332-x},
	language = {en},
	number = {1},
	urldate = {2025-06-02},
	journal = {Orphanet Journal of Rare Diseases},
	author = {Ghadessi, Mercedeh and Tang, Rui and Zhou, Joey and Liu, Rong and Wang, Chenkun and Toyoizumi, Kiichiro and Mei, Chaoqun and Zhang, Lixia and Deng, C. Q. and Beckman, Robert A.},
	month = dec,
	year = {2020},
	pages = {69},
}

@article{hobbs_hierarchical_2011,
	title = {Hierarchical {Commensurate} and {Power} {Prior} {Models} for {Adaptive} {Incorporation} of {Historical} {Information} in {Clinical} {Trials}},
	volume = {67},
	copyright = {http://doi.wiley.com/10.1002/tdm\_license\_1.1},
	issn = {0006341X},
	url = {https://academic.oup.com/biometrics/article/67/3/1047-1056/7380943},
	doi = {10.1111/j.1541-0420.2011.01564.x},
	language = {en},
	number = {3},
	urldate = {2025-06-02},
	journal = {Biometrics},
	author = {Hobbs, Brian P. and Carlin, Bradley P. and Mandrekar, Sumithra J. and Sargent, Daniel J.},
	month = sep,
	year = {2011},
	pages = {1047--1056},
}

@article{neuenschwander_note_2009,
	title = {A note on the power prior},
	volume = {28},
	copyright = {http://onlinelibrary.wiley.com/termsAndConditions\#vor},
	issn = {0277-6715, 1097-0258},
	url = {https://onlinelibrary.wiley.com/doi/10.1002/sim.3722},
	doi = {10.1002/sim.3722},
	language = {en},
	number = {28},
	urldate = {2025-06-02},
	journal = {Statistics in Medicine},
	author = {Neuenschwander, Beat and Branson, Michael and Spiegelhalter, David J.},
	month = dec,
	year = {2009},
	pages = {3562--3566},
}

@article{liao_prognostic_2025,
	title = {Prognostic adjustment with efficient estimators to unbiasedly leverage historical data in randomized trials},
	issn = {2194-573X, 1557-4679},
	url = {https://www.degruyterbrill.com/document/doi/10.1515/ijb-2024-0018/html},
	doi = {10.1515/ijb-2024-0018},
	language = {en},
	urldate = {2025-06-01},
	journal = {The International Journal of Biostatistics},
	author = {Liao, Lauren D. and Højbjerre-Frandsen, Emilie and Hubbard, Alan E. and Schuler, Alejandro},
	month = mar,
	year = {2025},
}

@article{schuler_increasing_2022,
	title = {Increasing the efficiency of randomized trial estimates via linear adjustment for a prognostic score},
	volume = {18},
	issn = {1557-4679},
	url = {https://www.degruyter.com/document/doi/10.1515/ijb-2021-0072/html},
	doi = {10.1515/ijb-2021-0072},
	language = {en},
	number = {2},
	urldate = {2025-06-01},
	journal = {The International Journal of Biostatistics},
	author = {Schuler, Alejandro and Walsh, David and Hall, Diana and Walsh, Jon and Fisher, Charles},
	month = dec,
	year = {2022},
	pages = {329--356},
}

@article{imbens_sensitivity_2003,
	title = {Sensitivity to {Exogeneity} {Assumptions} in {Program} {Evaluation}},
	volume = {93},
	issn = {0002-8282},
	url = {https://www.aeaweb.org/articles?id=10.1257/000282803321946921},
	doi = {10.1257/000282803321946921},
	language = {en},
	number = {2},
	urldate = {2025-05-21},
	journal = {American Economic Review},
	author = {Imbens, Guido W.},
	month = may,
	year = {2003},
	pages = {126--132},
}

@article{lim_minimizing_2018,
	title = {Minimizing {Patient} {Burden} {Through} the {Use} of {Historical} {Subject}-{Level} {Data} in {Innovative} {Confirmatory} {Clinical} {Trials}: {Review} of {Methods} and {Opportunities}},
	volume = {52},
	issn = {2168-4804},
	shorttitle = {Minimizing {Patient} {Burden} {Through} the {Use} of {Historical} {Subject}-{Level} {Data} in {Innovative} {Confirmatory} {Clinical} {Trials}},
	doi = {10.1177/2168479018778282},
	language = {eng},
	number = {5},
	journal = {Therapeutic Innovation \& Regulatory Science},
	author = {Lim, Jessica and Walley, Rosalind and Yuan, Jiacheng and Liu, Jeen and Dabral, Abhishek and Best, Nicky and Grieve, Andrew and Hampson, Lisa and Wolfram, Josephine and Woodward, Phil and Yong, Florence and Zhang, Xiang and Bowen, Ed},
	month = sep,
	year = {2018},
	pmid = {29909645},
	pages = {546--559},
}

@misc{chernozhukov_long_2024,
	title = {Long {Story} {Short}: {Omitted} {Variable} {Bias} in {Causal} {Machine} {Learning}},
	shorttitle = {Long {Story} {Short}},
	url = {http://arxiv.org/abs/2112.13398},
	doi = {10.48550/arXiv.2112.13398},
	urldate = {2025-01-29},
	publisher = {arXiv},
	author = {Chernozhukov, Victor and Cinelli, Carlos and Newey, Whitney and Sharma, Amit and Syrgkanis, Vasilis},
	month = may,
	year = {2024},
	note = {arXiv:2112.13398 [econ]},
}

@misc{noauthor_rare_2019,
	title = {Rare {Diseases}: {Natural} {History} {Studies} for {Drug} {Development}},
	url = {https://www.fda.gov/regulatory-information/search-fda-guidance-documents/rare-diseases-natural-history-studies-drug-development},
	month = mar,
	year = {2019},
}

@misc{ung_combining_2024,
	title = {Combining an experimental study with external data: study designs and identification strategies},
	shorttitle = {Combining an experimental study with external data},
	url = {http://arxiv.org/abs/2406.03302},
	doi = {10.48550/arXiv.2406.03302},
	urldate = {2025-02-19},
	publisher = {arXiv},
	author = {Ung, Lawson and Wang, Guanbo and Haneuse, Sebastien and Hernan, Miguel A. and Dahabreh, Issa J.},
	month = jun,
	year = {2024},
	note = {arXiv:2406.03302 [stat]},
}

@article{li_improving_2023,
	title = {Improving efficiency of inference in clinical trials with external control data},
	volume = {79},
	issn = {0006-341X, 1541-0420},
	url = {https://academic.oup.com/biometrics/article/79/1/394-403/7478074},
	doi = {10.1111/biom.13583},
	language = {en},
	number = {1},
	urldate = {2025-02-19},
	journal = {Biometrics},
	author = {Li, Xinyu and Miao, Wang and Lu, Fang and Zhou, Xiao‐Hua},
	month = mar,
	year = {2023},
	pages = {394--403},
}

\newpage
\appendix
\appendixpage
\section{Proofs}
\subsection{Identification Proof}
\label{identification-proof}
\begin{align*}
    E[Y_1|D=1] &= E[E[Y_1|X,D=1]]\\
    &= E[E[Y|A=1,X,D=1]] &&\text{(1)}\\
    \\[12pt]
    E[Y_0|D=1] &= E[E[Y_0|X,D=1]]\\
    &=E[E[Y|A=0, X, D=1]] &&\text{(1)}\\
    &= E[E[Y|A=0,X]] &&\text{(2)}
    \\[12pt]
    E[Y_1-Y_0|D=1]&= E[Y_1|D=1]-E[Y_0|D=1]\\
    &=E[E[Y|A=1,X,D=1]]-E[E[Y|A=0,X]]\\
    &=E[E[Y|A=1,X,D=1]-E[Y|A=0,X]]\\
\end{align*}
\begin{figure}[h]
    \centering
    \caption*{Assumptions: (1) Ignorability, (2) Mean Exchangeability of Controls}
\end{figure}

\subsection{Linear Functional of Trial-Specific ATE}
\label{rep}
\begin{align*}
    \Psi &= E[E[Y|A=1,X,D=1] - E[Y|A=0,X]]\\
    &= E[\frac{D}{q}(E[Y|A=1,X,D=1]-E[Y|A=0,X]]\\
    &= E[\frac{D}{q}(\mu(1,X)-\mu(0,X)]\\
    &= E[m(\mu)(O)]
\end{align*}
Where $m(\mu)(O)$ is defined as $\frac{D}{q}(\mu(1,X)-\mu(0,X))$ in the main text.
\newpage

\subsection{Riesz Representer of Trial-Specific ATE}
\label{rrproof}

\begin{align*}
    qE[E[Y|A=1,X,D=1]] &= E[\pi(X)E[Y|A=1,X,D=1]] \\
    &= E\left[\pi(X)\frac{D}{\pi(X)}E[Y|A=1,X,D]\right]\\
    &= E\left[\pi(X) \frac{D}{\pi(X)} \frac{A}{p(X)}E[Y|A,X,D]\right] &&\text{(1)}\\
    &= E\left[\frac{DA}{p(X)}E[Y|A,X,D]\right] \\
\end{align*}

\begin{align*}
    q[E[E[Y|A=0,X]]] &= E[\pi(X)E[Y|A=0,X]]\\
    &= E\left[\pi(X)E[Y|A=0,X,D=1]\frac{\pi(X)(1-p(X))}{\pi(X)(1-p(X))+(1-\pi(X))r(X)}\right.\\
    &\quad \left. +\pi(X)E[Y|A=0,X,D=0]\frac{(1-\pi(X))r(X)}{\pi(X)(1-p(X))+(1-\pi(X))r(X)}\right]  \tag{2}\\
    &= E \left[\pi(X)\frac{D(1-A)}{\pi(X)(1-p(X))}E[Y|A,X,D]\frac{\pi(X)(1-p(X))}{\pi(X)(1-p(X))+(1-\pi(X))r(X)}\right.\\
    &\quad \left. +\pi(X)\frac{(1-D)}{1-\pi(X)}E[Y|A,X,D]\frac{(1-\pi(X))r(X)}{\pi(X)(1-p(X))+(1-\pi(X))r(X)}\right] \tag{1}\\
    &= E[W(A,X,D)E[Y|A,X,D]]\\
\end{align*}

\begin{align*}
    E[E[Y|A=1,X,D=1]-E[Y|A=0,X]] &= E[E[Y|A=1,X,D=1]]-E[E[Y|A=0,X]]\\
    &= \frac{1}{q}E\left[\frac{DA}{p(X)}E[Y|A,X,D]\right]-\frac{1}{q}E[W(A,X,D)E[Y|A,X,D]]\\
    &= E\left[ \underbrace{\frac{1}{q} \left(\frac{DA}{p(X)} - W(A,X,D)\right)}_{\text{Riesz Representer}}  E[Y | A,X,D] \right]
\end{align*}

\begin{figure}[h]
    \centering
    \caption*{Assumptions: (1) Positivity, (2) Mean Exchangeability of Controls}
\end{figure}

Note that since we only use external data for control data, we also use our knowledge that $P(A=1|X,D=0)=0$ and $P(A=0|X,D=0)=1$ to construct weights. Further, in the second expansion of $q[E[E[Y|A=0,X]]$ we use optimal efficiency weights derived from the influence function of the target parameter (which combine propensity score weights and the variance ratio $r(X)$).
\section{Second-Order Remainders}

\subsection{Second-Order Remainder of Trial-Specific ATE}
\label{2tsate}

\begin{align*}
    R &= \hat \Psi -\Psi + E[\hat \phi(Y,A,X,D)]\\
    &= \hat \Psi - \Psi + E \left [\hat \alpha(A,X,D) (Y-\hat \mu(A,X)+m(\hat \mu)(O)-\frac{D}{\hat q}\hat \Psi \right]\\
    &= \hat \Psi -\Psi + E \left[\frac{1}{\hat q} \hat \beta(A,X,D)(Y-\hat \mu(A,X)+\frac{1}{\hat q} n(\hat \mu)(O) -\frac{D}{\hat q} \hat \Psi \right]\\
   &= \hat \Psi - \frac{1}{\hat q}\hat \Psi E D  +\frac{1}{\hat q} E \hat \beta(A,X,D) Y - \frac{1}{\hat q} E \hat \beta(A,X,D) \hat \mu(A,X) +\frac{1}{\hat q} E n(\hat \mu)(O) -\Psi \\
   &=\hat \Psi -\frac{q}{\hat q}\hat \Psi + \frac{1}{\hat q} E \hat \beta(A,X,D) \mu(A,X) -\frac{1}{\hat q} E \hat \beta(A,X,D) \hat \mu(A,X) + \frac{1}{\hat q} E \beta(A,X,D) \hat \mu(A,X) -\Psi\\
   &= \left(1-\frac{q}{\hat q}\right) \hat \Psi + \frac{1}{\hat q} E [-\hat \beta(A,X,D) \hat \mu(A,X) + \hat \beta(A,X,D) \mu(A,X) + \beta (A,X,D) \hat \mu(A,X)]-\Psi\\
   &= \left(1-\frac{q}{\hat q}\right) \hat \Psi + \frac{1}{\hat q} E [-\hat \beta(A,X,D) \hat \mu(A,X) + \hat \beta(A,X,D) \mu(A,X) + \beta(A,X,D) \hat \mu(A,X) \\
   &- \beta(A,X,D) \mu(A,X)] + \frac{1}{\hat q} E \beta (A,X,D) \mu (A,X) -\Psi\\
   % &= \left(1-\frac{q}{\hat q}\right) \hat \Psi - \frac{1}{\hat q} E [\hat \beta (A,X,D) \hat \mu (A,X) - \hat \beta(A,X,D) \mu(A,X) - \beta (A,X,D) \hat \mu (A,X) + \beta (A,X,D) \mu (A,X)] + \frac{1}{\hat q} E \beta (A,X,D) \mu (A,X) -\Psi\\
    &= \left(1-\frac{q}{\hat q}\right) \hat \Psi - \frac{1}{\hat q} E(\hat \beta(A,X,D) - \beta(A,X,D))(\hat \mu(A,X) -\mu(A,X)) + \frac{1}{\hat q} E \beta(A,X,D) \mu(A,X)-\Psi\\
    &= \left(1-\frac{q}{\hat q}\right) \hat \Psi - \frac{1}{\hat q} E(\hat \beta(A,X,D) - \beta(A,X,D))(\hat \mu(A,X) -\mu(A,X)) + \frac{1}{\hat q} E \beta(A,X,D) \mu(A,X)\\
    &-\frac{1}{q}E \beta(A,X,D) \mu(A,X)\\
    % &= \frac{1}{\hat q}(\hat q -q) \hat \Psi - \frac{1}{\hat q}\P (\hat \beta - \beta)(\hat \mu -\mu) + \frac{1}{\hat q} \P \beta \mu- \frac{1}{q} \P \beta \mu\\
    &= \frac{1}{\hat q}(\hat q -q) \hat \Psi - \frac{1}{\hat q} E (\hat \beta(A,X,D) - \beta(A,X,D))(\hat \mu(A,X) -\mu(A,X)) +  (\frac{1}{\hat q}-\frac{1}{q}) E \beta(A,X,D) \mu(A,X)\\
    &= \frac{1}{\hat q}(\hat q -q) \hat \Psi - \frac{1}{\hat q} E (\hat \beta(A,X,D) - \beta(A,X,D))(\hat \mu(A,X) -\mu(A,X) )+  \frac{1}{q}(\frac{q}{\hat q}-1) E\beta(A,X,D) \mu(A,X)\\
    % &= \frac{1}{\hat q}(\hat q -q) \hat \Psi - \frac{1}{\hat q}\P (\hat \beta - \beta)(\hat \mu -\mu) +  (\frac{q}{\hat q}-1) \Psi\\
    % &= \frac{1}{\hat q}(\hat q -q) \hat \Psi - \frac{1}{\hat q}\P (\hat \beta - \beta)(\hat \mu -\mu) +  \frac{1}{\hat q}(q-\hat q) \Psi\\
    &= \frac{1}{\hat q}(\hat q -q) \hat \Psi - \frac{1}{\hat q} E (\hat \beta(A,X,D) - \beta(A,X,D))(\hat \mu(A,X) -\mu(A,X)) -  \frac{1}{\hat q}(\hat q- q) \Psi\\
    &= \frac{1}{\hat q}(\hat q -q) (\hat \Psi - \Psi) - \frac{1}{\hat q}E (\hat \beta(A,X,D) - \beta(A,X,D))(\hat \mu(A,X) -\mu(A,X))\\
\end{align*}

Letting $n(\mu)(O)=D(\mu(1,X)-\mu(0,X))=qm(\mu)(O)$ and $\Psi=E(\frac{1}{q}\beta(A,X,D)\mu(A,X))=E(\alpha(A,X,D)\mu(A,X))$ under the Riesz Representation Theorem. Note that here we have manipulated two terms --- $m(\mu)(O)$ to $\frac{1}{q}n(\mu)(O)$ and $\alpha(A,X,D)$ to $\frac{1}{q} \beta$ --- to pull out the $\frac{1}{q}$ to explicitly highlight how the second-order remainder also depends on estimation of $q$.

Given the consistency of $\hat\Psi$ and the central limit theorem applied to $\hat q$, the first term is $o_P(n^{-1/2})$ as long as $q \ne 0$. The second term in the remainder is a typical ``doubly robust'' term: $\frac{1}{\hat q}E(\hat \beta(A,X,D)-\beta(A,X,D))(\hat \mu(A,X) -\mu(A,X))$. Typically for second-order convergence of the double robust term, we require that $\|\hat\mu -\mu\|\|\hat\beta-\beta\|$ is  $o_P(n^{-\frac{1}{2}})$ \cite{intro}. In our setting we must also require $q$ bounded away from 0. This is not an issue.

However, if $q$ is small we should expect a relatively larger impact of this term for fixed $n$, i.e. the asymptotics will not ``kick in'' until a relatively larger sample size. This behavior can be seen in figure 3 when external sample size is large but RCT sample size is small, leading to inflated type I error rates.

\section{Additional Figures}

\subsection{Type I Error Rate under Two Repetition Cross Fitting}
\label{ticf2}

\begin{figure}[H]
    \centering
    \includegraphics[width=1\textwidth]{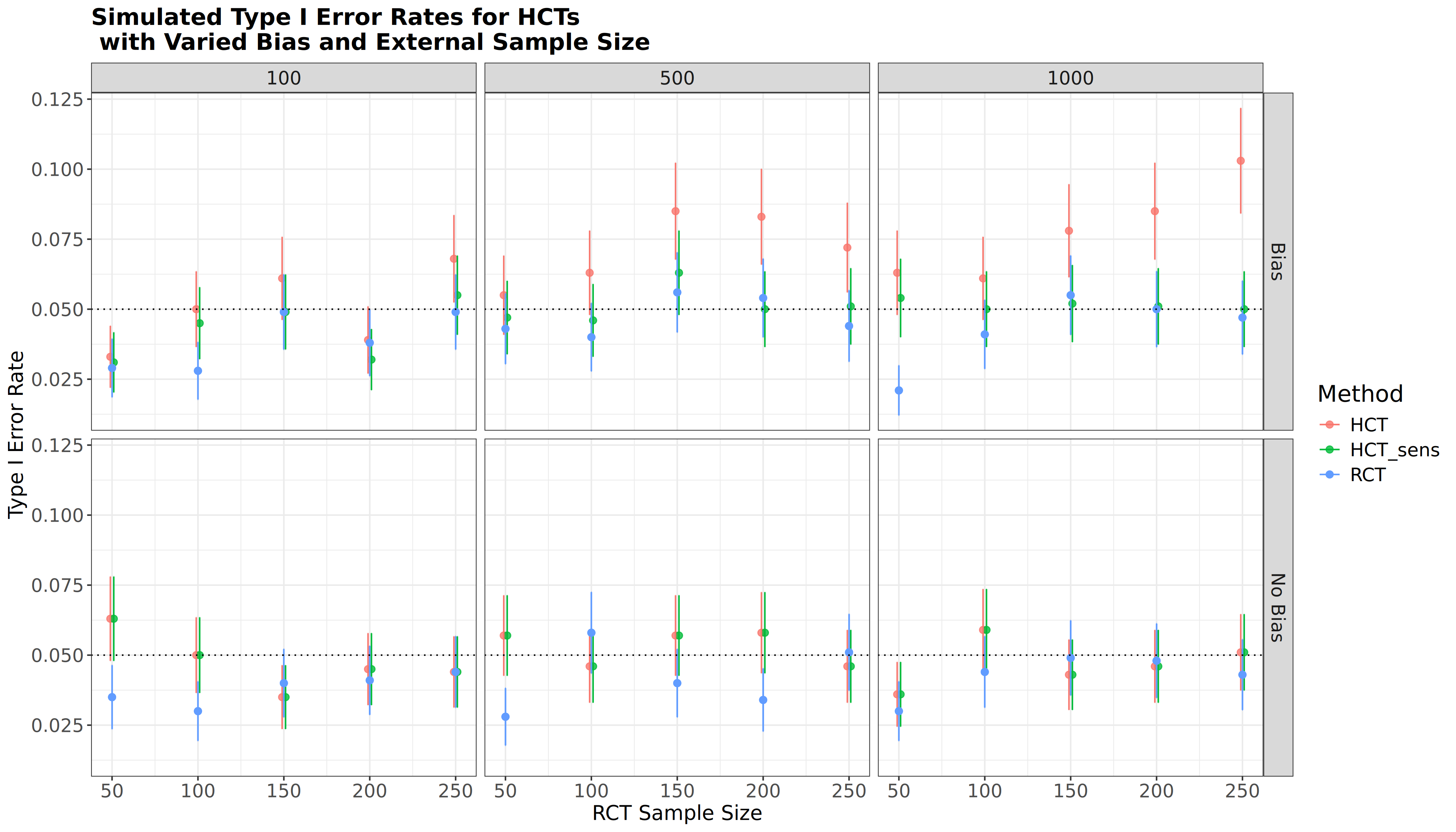}
    \caption{Type I error rates are compared across three trial setups: pure RCT, HCT, and HCT combined with the sensitivity analysis (shown in blue, red, and green respectively). Multiple scenarios are tested, varying the presence of bias and external sample sizes. From left to right: external sample size of 100, external sample size of 500, and external sample size of 1000. From top to bottom: bias present, no bias present. RCT sample sizes are varied with fixed external sample sizes.}
    \label{fig:cf2}
\end{figure}

\subsection{Contour Plot for Case Study under $\rho = 1$}
\label{app-rho1}

\begin{figure}[H]
    \centering
    \includegraphics[width=1\textwidth]{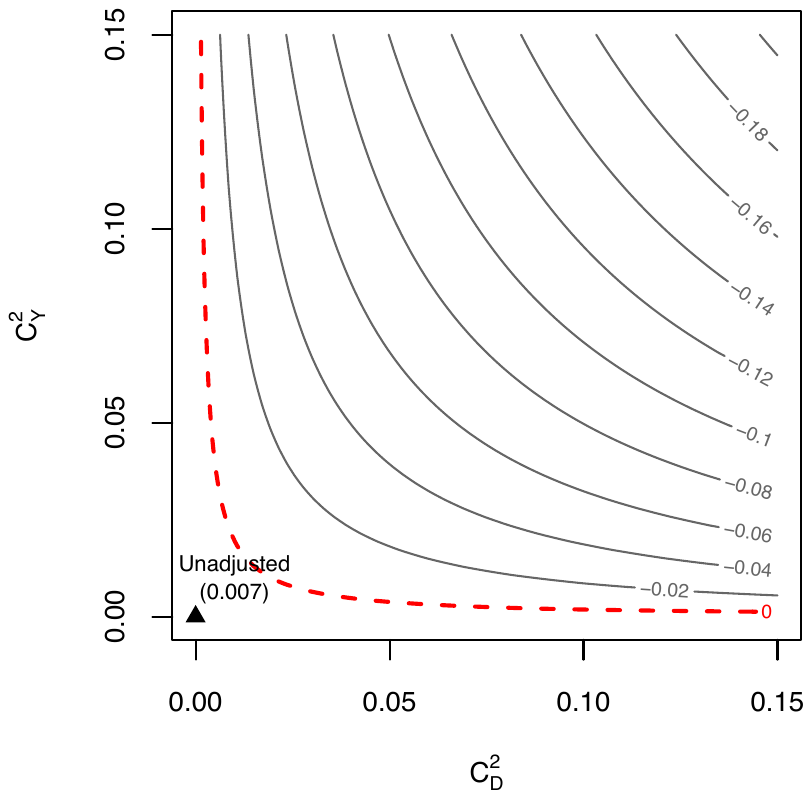}
    \caption{ Sensitivity contour plot for the HCT under $\rho=1$ for the most conservative estimates. The unadjusted estimate of the lower bound of the trial-specific average treatment effect is reported. The red dotted line represents the point at which results are no
longer statistically significant under combinations of $C^2_Y$ and $C^2_D$.}

\end{figure}

\end{document}